\documentclass[12pt]{article}
\usepackage{amssymb,epsfig}
\begin{document}

\title{
Quasi-periodic Green's functions of the Helmholtz 
and Laplace equations} 

\author{
Alexander Moroz\thanks{http://www.wave-scattering.com}
\\
Wave-scattering.com
} 

\maketitle

\begin{center}
{\large\sc abstract}
\end{center}
A classical problem of free-space Green's function $G_{0\Lambda}$ 
representations of the Helmholtz equation is studied in various
quasi-periodic cases, i.e., when an underlying periodicity 
is imposed in less dimensions than is the dimension of an embedding space. 
Exponentially convergent series for the free-space quasi-periodic $G_{0\Lambda}$
and for the expansion coefficients $D_{L}$ of $G_{0\Lambda}$ in the 
basis of regular (cylindrical in two dimensions and spherical in three 
dimension (3D)) waves, or lattice sums, are reviewed and new results 
for the case of a one-dimensional (1D) periodicity in 3D are derived. 
From a mathematical point of view, a derivation of exponentially 
convergent representations for Schl\"{o}milch series of cylindrical 
and spherical Hankel functions of any integer order is accomplished.
Exponentially convergent series for $G_{0\Lambda}$ and lattice sums $D_{L}$ 
hold for any value of the Bloch momentum and allow $G_{0\Lambda}$ to 
be efficiently evaluated also in the periodicity plane. The quasi-periodic 
Green's functions of the Laplace equation are obtained from the corresponding
representations of $G_{0\Lambda}$ of the Helmholtz equation by taking the limit 
of the wave vector magnitude going to zero. The derivation of relevant results 
in the case of a 1D periodicity in 3D highlights the common part which is 
universally applicable to any of remaining quasi-periodic cases. The results 
obtained can be useful for numerical solution of boundary integral equations 
for potential flows in fluid mechanics, remote sensing of periodic surfaces,
 periodic gratings, and infinite arrays of resonators coupled to a waveguide,
in many contexts of simulating systems 
of charged particles, in molecular dynamics, for the description of quasi-periodic 
arrays of point interactions in quantum mechanics, and in various ab-initio
first-principle multiple-scattering theories for the analysis of diffraction
of classical and quantum waves.

\maketitle  
\noindent \date{PACS numbers: 02.70.Pt; 03.65.Db; 03.65.Nk; 
34.20.-b; 42.25.Fx; 46.40.Cd; 47.15.km; 61.14.Hg}

 \thispagestyle{empty} 
\baselineskip 20pt 
\newpage 
\setcounter{page}{1}

\section{Introduction} 
Let $\Lambda$ be a $d_\Lambda$-dimensional simple (Bravais)  
periodic lattice embedded in a space of dimension 
$d\geq d_\Lambda$ (the condition of a simple lattice can 
easily be relaxed to an arbitrary periodic lattice by 
following recipes of Refs. \cite{Seg,HS,Bbl,Kam3,AmO}). 
Let  ${\cal H}_0^+$ stand for the 
 cylindrical ($H_{0}^{(1)}$) and spherical ($h_{0}^{(1)}$) Hankel functions 
in $d=2$  and $d=3$, respectively, ${\cal Y}_L$ be corresponding
angular momentum harmonics (cylindrical in $d=2$ and spherical in  $d=3$; see also 
Appendix \ref{app:sphr}) for properties of ${\cal Y}_{L}$), and
${\bf r}$ and ${\bf r}'$ be spatial points.  
The article is concerned with an efficient calculation
of the series
\begin{equation} 
 \sum_{{\bf r}_n\in\Lambda} {\cal H}_l^+(\sigma |{\bf r}-{\bf r}'+{\bf r}_n|)
 {\cal Y}_L^* (\hat{{\bf r}}_n)  e^{i{\bf k}\cdot{\bf r}_n},
\label{fsgfls} 
\end{equation} 
where the origin of coordinates 
is in the lattice, ${\bf k}$ is called the Bloch momentum, 
$\sigma =2\pi/\lambda$ denotes a wave vector magnitude ($\sigma$ is
 not necessarily equal to $|{\bf k}|$),
 $\lambda$ is a wavelength,  and 
$L$ is, in general, a multi-index of angular momentum numbers  
[e.g., $L=(lm)$ in three dimensions 
with $l\geq 0$ and $-l\leq m\leq l$].

In mathematical literature such a series
are known as Schl\"{o}milch series \cite{OXS,Wat,Tw}.
As first noted by Emersleben \cite{Em1}, in three dimensions (3D)
in the special case of ${\bf r}-{\bf r}'=0$ and
$l=\sigma=0$ the series reduce to 
{\em Epstein zeta functions} \cite{Ep,GlZ,BuC}.
A physical motivation to investigate such a series 
derives from a fact that, for $l=0$,
the series (\ref{fsgfls}) are prerequisite 
to determine a corresponding free-space (quasi-) periodic 
Green's function $G_{0\Lambda}$ of a scalar Helmholtz equation (see below).
For $l\neq 0$ and ${\bf r}-{\bf r}'=0$, with singular term being excluded,
the series (\ref{fsgfls}) then formally determine 
the lattice sums $D_{L}$, which are 
defined as the expansion coefficients $G_{0\Lambda}$
in the basis of cylindrical and spherical 
Bessel functions in 2D and 3D, respectively
[see Eqs. (\ref{dstrco}), (\ref{dstrcos}) below].

However, analytic closed expressions for such series are only known for $l=0$
in two particular cases in 3D:
(i) in the case of a one-dimensional (1D) periodicity \cite{Kar,Gla3,GHM}
and, when additionally $\sigma=0$ (the Laplace limit),
(ii) for a two-dimensional (2D) periodicity \cite{Gla3}.
Otherwise the summation in (\ref{fsgfls}) has to be performed
numerically. However, since  
\begin{equation}
{\cal H}_l^+(z) \sim c_h z^{-(d-1)/2}
\left(\frac{2}{\pi}\right)^{(d-1)/2} 
\exp\left\{ i\left[z -(d-1)\pi/4-l\pi/2\right]\right\}
\label{hlas}
\end{equation}
as $z\rightarrow\infty$ and  $-\pi<\arg{z}<2\pi$, 
where $c_h=1$ unless $d=2$, in which case
 $c_h=\sqrt{2/\pi}$
(see Eqs. (9.2.3) and (10.1.1) of Ref. \cite{AS}),
the series (\ref{fsgfls}) is
not absolutely convergent. 
Even if one assumes that $\sigma$ has an infinitesimally small positive 
imaginary part, and thereby establishing absolute convergence,
the convergence of the series in Eq. (\ref{fsgfls}) 
is notoriously slow, thereby rendering it
useless for practical applications.

The study of efficient techniques for the calculation of $G_{0\Lambda}$
 and lattice sums 
$D_{L}$ 
has a long history  \cite{Seg,HS,Kam3,Tw,Ign,Ew1,Ew2,Kam2,Kam1,Pen,OA,Be}
and the topic has roots and branches in widely differing areas
of chemistry, physics, and mathematics. 
Despite that it still continues to be a perennial research subject 
\cite{AmO,VYS,Mol,Moc,MN,Lin,Lnt,EP,GJ,NMcP,SC,YYo,PMBG,CST}. 
However, quite often this happens merely because
researches  in widely differing areas
of chemistry, physics, and mathematics are not aware of the techniques
and results developed in connection with the so-called 
Korringa-Kohn-Rostoker (KKR)
 \cite{Seg,HS,OA,Be} and
layer Korringa-Kohn-Rostoker (LKKR) theories 
either for quantum (electron) waves within low-energy 
electron diffraction (LEED) theory
\cite{Kam3,Kam2,Kam1,Pen} or for various classical (acoustic, elastic,
electromagnetic, water) waves \cite{AmO,Mol,Moc}. 
Therein {\em exponentially convergent} series for $G_{0\Lambda}$
 and lattice sums  $D_{L}$ have been derived
for a number of cases. In a {\em periodic case} ($d_\Lambda=d$),
efficient 
computational schemes for $G_{0\Lambda}$ and lattice sums  $D_{L}$ 
have been provided more than thirty years ago \cite{HS,OA}.
The  quasi-periodic 
case $d_\Lambda=2$ and $d=3$ has been investigated 
in detail in a series of 
articles by Kambe almost forty years ago \cite{Kam3,Kam2,Kam1}. 
The case $d_\Lambda=1$ and $d=2$
has been dealt with only relatively recently in Refs. \cite{AmO,Mol}. 
Surprisingly enough, the 
case $d_\Lambda=1$ and $d=3$ has not been studied in full detail yet, although
it may provide an 
efficient description of ``wave integrated circuits". 

Therefore, in the following we shall focus on the so-called
{\em quasi-periodic}, or layer, case, when the underlying lattice
 $\Lambda$ 
is of lower dimensionality than the embedding space ($d_\Lambda<d$), 
and, in particular, on the $d_\Lambda=1,\, d=3$ case.
There are many physical problem which would profit 
from an efficient computational scheme
for $G_{0\Lambda}$ and lattice sums $D_{L}$  in the case of 1D periodicity in 3D.
For instance, a Green's function representing a point source and 
satisfying the respective 
von Neumann and Dirichlet boundary condition on 
a flow channel walls in fluid mechanics for the flow between parallel planes
can be written
as a sum and difference of  $G_{0\Lambda}$ corresponding to 1D periodicity in 3D 
taken at two different spatial points \cite{Lnt}.
(For the flow in a rectangular channel, the relevant $G_{0\Lambda}$ would then
correspond to 2D periodicity in 3D \cite{Lnt}.) 
Another problem involves
 infinite arrays of resonators coupled to a waveguide which by itself
may have a plethora of photonics applications \cite{ChU}.
Moreover, recent progress in nanotechnology made it possible
to fabricate 1D chains of metal nanoparticles \cite{BHA,MKA}
and dielectric microparticles \cite{AFA}.
Additionally, understanding of linear periodic arrays of lossless spheres 
is germane for the qualitative description of a finite-lengths 
periodic arrays of small antennas \cite{SYa}.
In a linear chain of  spherical metal nanoparticles 
 light can be transmitted by electrodynamic interparticle  coupling
resulting in a subwavelength-sized light guide
\cite{BHA,MKA,QLK,PaS,WeF}.
So far, infinite 1D linear chains of particles have only been 
investigated within the one particle theory framework of 
Schr\"{o}dinger equation for the description of polymers \cite{GHM},
in the electrostatic limit within the framework of the
Laplace equation \cite{PaS}, or in the dipole approximation
\cite{SYa,WeF,GAu1}. Moreover, the energy operator
in the one-particle theory of periodic point (zero-range) interactions 
is constructed in terms of an auxiliary operator 
which corresponds essentially to the operator of multiplication by 
$D_{00}$ ($\gamma$ function of Karpeshina \cite{Kar,Krp}).

In the following, exponentially convergent series for the 
free-space quasi-periodic $G_{0\Lambda}$
and for the lattice sums  $D_{L}$ are reviewed and new results 
for the case of a 1D periodicity in 3D are derived
[see Eqs. (\ref{dl1f1in3}), (\ref{dl2f12in3}), and (\ref{dl3f1in3}) below].
The derivation of relevant results 
for the case of a 1D periodicity in 3D is performed in such a way
that the common part which is 
universally applicable to any of the remaining quasi-periodic cases
is highlighted. Thereby a link with earlier results by Kambe
\cite{Kam3,Kam2} for a 2D periodicity in 3D can easily be established
and the proof of results for a 1D periodicity in 2D announced
in Ref. \cite{Mol} can easily be carried out.

\subsection{The outline of the article} 
The article is organized as follows.
Sec. \ref{sec:notd} introduces notation, provides some necessary
definitions and gives an overview of some of the
 problems requiring the knowledge
$G_{0\Lambda}$ and lattice sums.
In Sec. \ref{sec:dual},  $G_{0\Lambda}$ is  
expressed as an exponentially convergent sum over a dual lattice 
$\Lambda^*$. Such a dual representation 
of $G_{0\Lambda}$ is then a 
starting point in the derivation of a corresponding exponentially 
convergent Ewald representation of 
$G_{0\Lambda}$ in Sec. \ref{sec:ew}. As in the bulk case, 
a derivation of the Ewald representation invokes a suitable 
integral representation of 
Hankel functions and a Jacobi identity (see Appendix \ref{app:jacobi} below).
In addition, following Kambe \cite{Kam1}, an analytic  continuation 
procedure is employed, which is
analogous to finding the values of the Riemann $\zeta$-function outside the domain 
of absolute convergence of a defining series
(see Ref. \cite{WWat}, p. 273).
The Ewald representation of $G_{0\Lambda}$, which is a hybrid sum
over both $\Lambda$
and dual lattice $\Lambda^*$,
converges uniformly and absolutely over bounded sets 
of ${\bf R}$. Unlike the dual representation, 
the Ewald representation can also 
be efficiently evaluated in the periodicity plane (provided that it remains 
${\bf R}\not\in\Lambda$).

Exponentially convergent series for the 
lattice sums $D_{L}$ in the case of a 1D periodicity in 3D are derived
in Sec. \ref{sec:dlcalc}.
 Eqs. (\ref{dl1f1in3}), (\ref{dl2f12in3}), (\ref{dl3f1in3})
of the section are the main new results of this paper.

In Sec. \ref{sec:pois} the quasi-periodic Green's function of the Laplace 
equation are obtained from that of the Helmholtz equation by taking the limit 
$\sigma \rightarrow 0$. We then end up with discussion (Sec. \ref{sec:disc}) 
and summary and conclusions (Sec. \ref{sec:concl}).

To make this paper as readable as possible, several technical 
arguments have been relegated to a number of appendices.
Appendix \ref{app:hintr} summarizes relevant integral representations of 
${\cal H}^+_l$ and Appendix \ref{app:sphr} lists relevant
properties of harmonics ${\cal Y}_L$. Some useful properties of 
free-space scattering Green's function are collected in 
Appendix \ref{app:fsgf}, whereas  
Appendix \ref{app:jacobi} shows several forms of Jacobi identities.
In Appendix \ref{app:fsqpgf} 
the dual representation of the quasi-periodic Green's functions 
is derived by
following original path of McRae \cite{McRaer} 
for a 2D periodicity in 3D, i.e., by applying 
an Ewald integral representation 
and the generalized Jacobi identity.
Some of the general properties of free-space quasi-periodic  
Greens functions and the lattice sums are 
outlined in Appendix \ref{sec:gfgenprlc}.
Alternative definitions of lattice sums and structure constants 
are then summarized in Appendix \ref{app:alter}.

\section{Notation and definitions: $G_{0\Lambda}$ and lattice sums} 
\label{sec:notd}

\subsection{$G_{0\Lambda}$}   
A corresponding free-space (quasi-) periodic 
Green's function $G_{0\Lambda}$
of a scalar Helmholtz equation is defined by an image-like series
\begin{equation}
G_{0\Lambda}(\sigma,{\bf k},{\bf R})
=\sum_{{\bf r}_n\in\Lambda} G_0^+ (\sigma,{\bf R}+{\bf r}_n) 
e^{-i{\bf k}\cdot{\bf r}_n}
=\sum_{{\bf r}_n\in\Lambda} G_0^+ (\sigma,{\bf R}-{\bf r}_n) 
e^{i{\bf k}\cdot{\bf r}_n},
\label{lgrfdf}
\end{equation}
where the origin of coordinates is in the lattice. Here
$G_0^+$ is the free-space scattering, or retarded, Green's function
of a scalar Helmholtz equation in $d$ dimensions,
\begin{equation} 
\left[\Delta + \sigma^2\right] G_0(\sigma,{\bf r},{\bf r}')
  =\delta({\bf r}-{\bf r}'), 
\label{helmeq} 
\end{equation} 
$\Delta$ being  a corresponding Laplace operator,
which is represented for large $R=|{\bf R}|=|{\bf r}-{\bf r}'|$ 
by outgoing waves
(i.e. satisfies Sommerfeld radiation condition).
Since $G_0^+$ is only a function of 
${\bf R}={\bf r}-{\bf r}'$, the functional dependence of $G_0^+$
has been written as $G_0^+ (\sigma,{\bf R})$
in Eq. (\ref{lgrfdf}). 

$G_0^+$ is  proportional to an appropriate Hankel function of zero order \cite{AS}
\begin{equation} 
G_0^+(\sigma,{\bf R}) = \lim_{\epsilon\rightarrow 0_+} \frac{1}{(2\pi)^d} 
\int \frac{e^{i{\bf k}\cdot{\bf R}}}
 {\sigma^2-k^2+ i\epsilon}\,d{\bf k} 
= -i \frac{\pi}{2}\, \frac{A}{(2\pi)^d}\, \sigma^{d-2} 
   {\cal H}_0^+(\sigma R) 
\label{fsgf} 
\end{equation} 
(see also Appendix \ref{app:fsgf}). 
Here $k=|{\bf k}|$ and $A$ is the surface of a unit sphere in $d$-dimensions 
[see Eq. (\ref{Adef}) below].
(With ${\cal H}_0(z) = e^{i z}$ as a one-dimensional (1D) 
analog of the Hankel function \cite{But}
Eq. (\ref{fsgf})  becomes also valid for $d=1$, in which case $A=2$
(unit ``sphere" in 1D consists of two points).)
Consequently,  upon
substituting (\ref{fsgf}) into (\ref{lgrfdf}) one arrives 
up to a proportionality factor 
to the special case of series (\ref{fsgfls}) for $l=0$.

A scalar Helmholtz equation 
is employed for the description
of various waves arising in acoustics, mechanics, fluid dynamics,
electromagnetism, and quantum mechanics \cite{GSt}.
An important class of problems which requires an 
efficient calculation of $G_{0\Lambda}$ arises in connection with 
remote sensing of periodic surfaces \cite{VYS},
numerical solution of boundary integral equations 
for potential flows in fluid mechanics \cite{MN,Lin,Lnt,EP}, 
 periodic gratings,
and infinite arrays of resonators coupled to a waveguide \cite{ChU},
in descriptions of dipolar fields in simulated liquid-vapour interfaces
\cite{Lek1}, in many contexts 
of simulating systems of charged particles, such as crystal binding 
and lattice vibrations \cite{Em1,GlZ}, Madelung constant \cite{GlZ,BuC}, 
in various problems in molecular dynamics
and Monte Carlo simulations of particles interacting by long-range
 Coulomb forces \cite{GJ,TB}, wherein
periodic boundary conditions are usually imposed in
order to avoid the boundary effects.

\subsection{Lattice sums} 
Within a primitive cell of $\Lambda$, the respective
Green's functions $G_0^+$ and $G_{0\Lambda}$  
only differ up to boundary conditions and their 
respective singular parts are  identical. 
For the scalar Helmholtz equation, the singular, or principal-value, 
part of $G_0^+$ is 
\begin{equation}  
G_0^p(\sigma,{\bf R})=\mbox{Re}\, G_0^+ (\sigma,{\bf R})
=\left\{  
\begin{array}{cc}  
 N_0 (\sigma R)/4,  & 2D\\  
-\cos (\sigma R)/(4\pi R), & 3D  
\end{array}\right.  
\label{singp}  
\end{equation}  
where $N_0$ is the cylindrical Neumann function \cite{AS}.  
Therefore, the difference 
\begin{equation}
D_\Lambda(\sigma,{\bf k},{\bf R})
 = G_{0\Lambda}(\sigma,{\bf k},{\bf R}) - G_0^p(\sigma,{\bf R})
\label{dlmbd}
\end{equation}
is regular for ${\bf R} \rightarrow 0$  
and can be expanded in terms of   
regular [cylindrical in 2D, spherical in 3D] waves 
\cite{HS,Kam3,Kam2,OA,Mol,Moc,KR}  
\begin{equation}  
D_\Lambda(\sigma,{\bf k},{\bf R}) = \sum_{L} D_{L}(\sigma,{\bf k}) 
{\cal J}_{l}(\sigma  R) {\cal Y}_L(\hat{{\bf R}}).
\label{dstrco} 
\end{equation}
Here the symbol 
${\cal J}_{l}$  stands for cylindrical and  spherical 
Bessel functions in 2D and 3D, respectively.
The expansion coefficients $D_{L}(\sigma,{\bf k})$ 
introduced by Eq. (\ref{dstrco}) are then the sought {\em lattice sums}.
The choice of what to subtract off in Eq. (\ref{dlmbd}) is somewhat arbitrary
and other choices will lead to slight amended expressions 
for the lattice sums $D_{L}$ 
(see also Appendix \ref{app:alter}). The present choice
goes back to Kohn and Rostoker \cite{KR} and has been 
adopted by Ham and Segall \cite{HS}, Kambe \cite{Kam3,Kam2},
 Pendry \cite{Pen}, and others \cite{Mol,Moc}.

The series  (\ref{fsgfls}) for $l\neq 0$ then formally determine 
the lattice sums $D_{L}$.  
Indeed, the free space Green's function is known to possess a
partial wave expansion,
\begin{equation} 
G_0^+ (\sigma,{\bf r},{\bf r}') = -i AC \sum_L {\cal J}_l(\sigma r_<)  
{\cal H}_l^{+}(\sigma r_>) {\cal Y}_L(\hat{{\bf r}}_<) {\cal Y}_L^*(\hat{{\bf r}}_>), 
\label{gfpwex} 
\end{equation} 
where $r_>$ ($r_<$) is the larger (smaller) of 
the $|{\bf r}|$ and $|{\bf r}'|$. The  numerical constant 
C [see Eq. (\ref{ACfac}) or Eq. (\ref{CArel}) below]
 is basically the prefactor 
in Eq. (\ref{fsgf}). Note that 
\begin{equation} 
AC= \frac{\pi}{2}\, \frac{A^2}{(2\pi)^d}\, \sigma^{d-2} 
=\left\{ 
\begin{array}{ll} 
\frac{1}{\sigma}, & 1D\\ 
\frac{\pi}{2}, & 2D\\ 
\sigma, & 3D. 
\end{array} \right. 
\label{ACfac} 
\end{equation} 
When the
partial wave expansion  is substituted for 
$G_0^+$ in the series (\ref{lgrfdf}) for $G_{0\Lambda}$,
then, according to Eqs. (\ref{dlmbd}), (\ref{dstrco}), one has
\begin{equation}  
 D_{L}(\sigma,{\bf k}) = -iCA^{1/2}\delta_{L0} - i AC
\sum_{{\bf r}_n\in\Lambda}{}^{'}  {\cal H}_{l}^+(\sigma r_n) 
{\cal Y}_L^*(\hat{{\bf r}}_n) e^{i{\bf k}\cdot{\bf r}_n},
\label{dstrcos} 
\end{equation} 
where a prime on summation sign will  here and hereafter indicate
that the term ${\bf r}_n=0$ is omitted from the sum.
Note that the series is, up to a constant term and a proportionality factor, 
the special case of series (\ref{fsgfls}) for $l\neq 0$ and ${\bf R}=0$.

It is worthwhile to point out that
 one is often rather interested in the lattice sums 
$D_{L}$ than at $G_{0\Lambda}$ itself. For instance, 
since $D_{L}{}$'s do not depend on ${\bf R}$, 
$G_{0\Lambda}$ can be evaluated using  Eqs. (\ref{dlmbd})
and (\ref{dstrco}) at any observation point
with the same set of the lattice sums.
 This can in turn significantly speed-up
numerical solutions of various boundary-value problems.
Knowledge of the lattice sums $D_{L}$
is a key to 
efficient numerical analysis of various ab-initio
first-principle multiple-scattering 
problems, such as band structure calculation
 within Korringa-Kohn-Rostoker (KKR) theories 
\cite{Seg,HS,KR,OA,Kor,F5,RGNb,Go,WZY,Mo,MS,LM,AM2,KE,AMprb},
and diffraction  problems by periodic structures, or gratings,
using the so-called
layer KKR (LKKR) theories or equivalents thereof, 
either for quantum (electron) waves within low-energy 
electron diffraction (LEED) theory
\cite{Kam3,Kam2,Pen,Tong,McLC} or, for various classical (acoustic, elastic,
electromagnetic, water) waves 
\cite{AmO,Mol,AMprb,Oh,Mod,SKM,Mays,YSM,PSM,PTK}. 
Lattice sums also arise in quantizing
classically ergodic systems such as Sinai's billiard \cite{Be} and its 
various electromagnetic analogs \cite{Be,QCh}.
Moreover, the spectrum within
 the one-particle theory of periodic point (zero-range) interactions  is 
determined as the set of those $z$ which, for a given
${\bf k}_\parallel$ and $\alpha$, satisfy an implicit equation \cite{Kar,Krp}
\begin{equation}
 D_{00}(i \sqrt{-z} ,{\bf k}_\parallel) = \tilde{\alpha}.
\label{zps}
\end{equation}
The spectral parameter $\tilde{\alpha}$ here is a boundary-condition 
parameter which determines the asymptotic of eigenfunctions 
at the lattice points and is the
same for all the eigenfunctions \cite{Kar,Krp,AGH}.

\section{Dual representations of quasi-periodic $G_{0\Lambda}$}
\label{sec:dual} 
Let $\Lambda^*$ be a corresponding dual 
(momentum) lattice, i.e., for any ${\bf r}_n\in \Lambda$ and  
${\bf k}_s\in\Lambda^*$ one has ${\bf r}_n\cdot{\bf k}_s=2 \pi N$,  
 where $N$ is an integer. 
It is well known that $G_{0\Lambda}$ has an alternative
representation as a sum over the dual lattice $\Lambda^*$.
For example, in the bulk case, i.e., when $d_\Lambda=d$,  
the dual sum representation of $G_{0\Lambda}$ is 
\begin{equation} 
G_{0\Lambda}(\sigma,{\bf k},{\bf R}) = \frac{1}{v_0} \sum_{{\bf k}_s\in\Lambda^*} 
\frac{e^{i({\bf k}+{\bf k}_s)\cdot{\bf R}}}
           {\sigma^2 - ({\bf k}+{\bf k}_s)^2 }
=  \sum_{{\bf k}_s\in\Lambda^*} 
\frac{\psi({\bf k}+{\bf k}_s,{\bf r}) \overline{\psi} ({\bf k}+{\bf k}_s,{\bf r}')}
           {\sigma^2 - ({\bf k}+{\bf k}_s)^2},
\label{dualbulk} 
\end{equation}  
with eigenfunctions 
\begin{equation}
\psi({\bf k}+{\bf k}_s,{\bf r})=\frac{1}{\sqrt{v_0}}\, 
e^{i({\bf k}+{\bf k}_s)\cdot{\bf r}}
\label{eigfnc}
\end{equation}
 normalized to unity in the fundamental (Wigner-Seitz) domain,
\begin{equation}
\int_{WS}\bar{\psi}({\bf k}+{\bf k}_s,{\bf r}) 
\overline{\psi}_({\bf k}+{\bf k}_{s'},{\bf r})\,d{\bf r}=\delta_{ss'},
\label{eigfncnm}
\end{equation}
$v_0$ being the volume of a unit-cell of $\Lambda$.
A dual representation is sometime called
an eigenfunction expansion of $G_{0\Lambda}$.
Often the respective representations 
(\ref{lgrfdf}) and (\ref{dualbulk})  
of $G_{0\Lambda}$ are also called the {\em spatial-domain} 
and {\em spectral-domain} forms,
respectively \cite{MN,Lin,EP,NMcP,YYo}).

In the following, the dual sum representation of $G_{0\Lambda}$ in
the quasi-periodic case of 1D periodicity in 3D will be derived
and its convergence properties will be discussed.
However, before proceeding any further, it turns out expedient
to provide some supplementary geometrical definitions
which are up to minor variations
adopted, for instance, within LEED and LKKR theories
\cite{Kam3,Kam2,Pen,Mol,Moc,Tong,McLC,Oh,Mod,SKM,YSM,PSM,PTK}.

\subsection{Supplementary geometrical definitions} 
In the quasi-periodic case we define the respective parallel 
and perpendicular components ${\bf r}_\parallel$ and  ${\bf r}_\perp$   
of a given vector ${\bf r}={\bf r}_\parallel + {\bf r}_\perp$ 
with respect to the $d_\Lambda$-dimensional 
plane containing the Bravais lattice 
$\Lambda$ (i.e., line for $d_\Lambda=1$ and surface for $d_\Lambda=2$)
 and its normal ${\bf n}$ , 
respectively. Then
for any ${\bf r}_n\in\Lambda$, 
\begin{equation} 
{\bf r}\cdot{\bf r}_n={\bf r}_\parallel\cdot{\bf r}_n,   
\hspace*{1.5cm}{\bf r}_\perp\cdot{\bf r}_n\equiv 0. 
\nonumber
\end{equation}
The respective projections ${\bf k}_\parallel$
and ${\bf k}_\perp$ of wave vector ${\bf k}={\bf k}_\parallel + {\bf k}_\perp$
are then defined in like manner with respect to $\Lambda^*$.
Obviously, in a quasi-periodic case the quantities 
$G_{0\Lambda}(\sigma,{\bf k},{\bf R})$, $D_\Lambda(\sigma,{\bf k},{\bf R})$
and $D_{L}(\sigma,{\bf k})$ entering Eqs. 
(\ref{lgrfdf}), (\ref{dlmbd}), (\ref{dstrco}),
(\ref{dstrcos}) are only functions of ${\bf k}_\parallel$.
Therefore, in the quasi-periodic case, i.e., 
if the underlying lattice $\Lambda$ is of lower dimensionality than 
the embedding space ($d_\Lambda<d$), it is more appropriate to call
merely the projection ${\bf k}_\parallel$ as the Bloch momentum.  

\begin{figure}[tbp]
\begin{center}
\epsfig{file=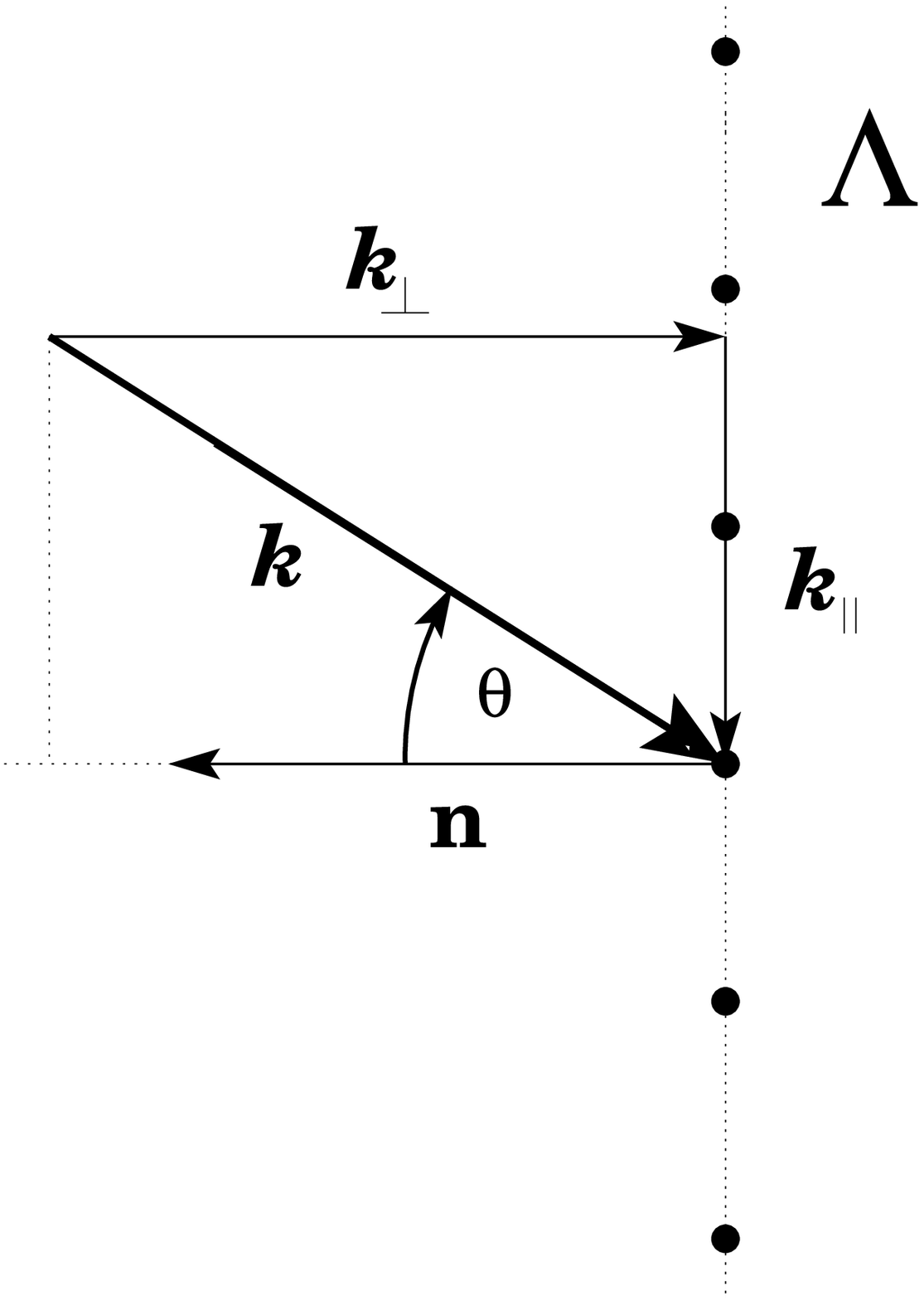,width=12cm,clip=0,angle=0}
\end{center}
\caption{
}
\label{fgpwinc}
\end{figure}

A plane wave $e^{i{\bf k}\cdot{\bf r}}$ incident on a 
scattering plane of identical scatterers arranged regularly 
on $\Lambda$ would (see Fig. \ref{fgpwinc}), in general, be diffracted (transmitted) 
to a wave with a wave vector ${\bf K}_n^-$ (${\bf K}_n^+$), 
where ${\bf K}_{n}^\pm = \left({\bf k}_\parallel+{\bf k}_n,  
K_{\perp n}^\pm \right)$, 
\begin{equation}
K_{\perp n}^\pm = \pm K_{\perp n}=\left\{
\begin{array}{cc}
\pm [\sigma^2-|{\bf k}_\parallel+{\bf k}_n|^2]^{1/2},
      &\sigma^2 \geq |{\bf k}_\parallel+{\bf k}_n|^2\\
\pm i[|{\bf k}_\parallel+{\bf k}_n|^2-\sigma^2]^{1/2},
      &\sigma^2<|{\bf k}_\parallel+{\bf k}_n|^2,
\end{array}\right.
\label{Kkn}
\end{equation}
${\bf k}_n\in \Lambda^*$, and $|{\bf k}|=|{\bf K}_{n}^\pm|=\sigma$. 
Here $K_{\perp n}$ is indicated as a scalar, which is definitely
true for $d-d_\Lambda=1$. In the case of a 1D periodic chain in 3D ($d-d_\Lambda=2$)
$K_{\perp n}$ will be taken as the projection of ${\bf K}_{\perp n}$ 
on the plane spanned by the wave vectors of incident and diffracted
beams.
In the above definition, the projection  
${\bf K}_{\parallel n}={\bf k}_\parallel+{\bf k}_n$ 
is real but the normal projection $K_{\perp n}$ 
can be either real or imaginary. In the case of real 
$K_{\perp n}$ we speak of a {\em propagating wave}, and 
in the case of imaginary $K_{\perp n}$ of an {\em evanescent wave}. 

In the present case the scatterers
are absent. Nevertheless, it turns out expedient to
define wave vectors ${\bf K}_{n}^\pm$ and the respective projections
${\bf K}_{\parallel n}$ and $K_{\perp n}^\pm$ even in the free-space case.

\subsection{Resulting series}
In order to establish a dual representation of a quasi-periodic 
$G_{0\Lambda}$ 
for $d-d_\Lambda=2$, i.e. a 1D periodicity along the $x$-axis in 3D,
one first substitutes the integral representation (\ref{fsgf})
of $G_0^+$ into defining equation (\ref{lgrfdf})
of a free-space quasi-periodic Green's function $G_{0\Lambda}$.
Then the Poisson formula 
\begin{equation}
\sum_{{\bf r}_n\in\Lambda} e^{i({\bf q}_\parallel - 
    {\bf k}_\parallel)\cdot{\bf r}_{\parallel n}}
= \frac{(2\pi)^{d_\Lambda}}{v_0} \sum_{{\bf k}_n\in\Lambda^*}
\delta({\bf q}_\parallel -{\bf k}_\parallel - {\bf k}_n)
\label{poisd}
\end{equation}
is applied resulting in
\begin{equation}
G_{0\Lambda}(\sigma,{\bf k}_\parallel,{\bf R})
= \frac{(2\pi)^{d_\Lambda-d} }{v_0}
 \sum_{{\bf k}_n\in\Lambda^*} \int 
\frac{e^{i{\bf q}_\perp\cdot 
{\bf R}_\perp + i({\bf k}_\parallel+{\bf k}_n)\cdot {\bf R}_\parallel} } 
{\sigma^2-{\bf q}_\perp^2- |{\bf k}_\parallel+{\bf k}_n|^2 + i\epsilon }\, d {\bf q}_\perp.
\label{lgrfdf13}
\end{equation}
Now the 2D plane-wave expansion (\ref{gplw}) is applied to  
$e^{i{\bf q}_\perp\cdot{\bf R}_\perp}$.
Using the orthonormality of cylindrical harmonics
$Y_l=e^{il\phi} /\sqrt{2\pi} $ 
one finds [Eq. (\ref{sphryps0})] 
\begin{equation}
\int_0^{2\pi}  Y_l(\phi)\,d\phi =\sqrt{2\pi}\, \delta_{l0}.
\nonumber
\end{equation}
Therefore, integration in the integral representation (\ref{lgrfdf13})
of $G_{0\Lambda}$ over ${\bf q}_\perp$ results in
\begin{eqnarray} 
G_{0\Lambda} (\sigma,{\bf k}_\parallel,{\bf R}) &=&  \frac{(2\pi)^{d_\Lambda-d+1} }{v_0}
 \sum_{{\bf k}_n\in\Lambda^*} e^{i({\bf k}_\parallel+{\bf k}_n)\cdot {\bf R}_\parallel} 
\int_0^\infty  \frac{ q_\perp J_0(q_\perp |R_\perp|)
}{\sigma^2-q_\perp^2- |{\bf k}_\parallel+{\bf k}_n|^2 + i\epsilon }\, d q_\perp 
\nonumber\\
&=& -\frac{i}{4v_0} \sum_{{\bf k}_n\in\Lambda^*} 
e^{i({\bf k}_\parallel+{\bf k}_n)\cdot {\bf R}_\parallel} 
 H_0(K_{\perp n} |R_\perp|).
\label{31dual}
\end{eqnarray}
Here in going from the first to second equality the integral identity
(\ref{int0}) for 2D case has been applied.
Thereby the sum over ${\bf r}_n\in\Lambda$ has been transformed into a sum
over ${\bf k}_n\in\Lambda^*$ resulting in the so-called dual representation 
(spectral domain form) of $G_{0\Lambda}$.

\subsubsection{Complementary cases}
For completeness, in the case of codimension one ($d-d_\Lambda=1$), 
one first performs a partial integral over ${\bf R}_\perp$ 
in the integral representation (\ref{fsgf}) of
 free-space Green's function.
This amounts to picking up a residue of a contour integral 
in the complex plane according to Cauchy theorem resulting in 
\begin{equation} 
G_0^+ (\sigma,{\bf R}) = - \frac{\pi i}{(2\pi)^d}
 \int \frac{e^{i{\bf q}_\parallel\cdot{\bf R}_\parallel
+i \sqrt{\sigma^2-{\bf q}_\parallel^2}|R_\perp|}
}{\sqrt{\sigma^2-{\bf q}_\parallel^2}}\,d {\bf q}_\parallel .
\label{fspigf} 
\end{equation} 
Substituting the integral 
representation of $G_0^+ (\sigma,{\bf R})$ into defining 
equation (\ref{lgrfdf}) of a free-space quasi-periodic Green's 
function $G_{0\Lambda}$ then results in
\begin{eqnarray}
G_{0\Lambda}(\sigma,{\bf k}_\parallel,{\bf R})
&=&- \frac{\pi i}{(2\pi)^d} 
\int \frac{e^{i{\bf q}_\parallel\cdot {\bf R}_\parallel
+ i \sqrt{\sigma^2-{\bf q}_\parallel^2}|R_\perp|}
}{\sqrt{\sigma^2-{\bf q}_\parallel^2}}\,
\left[\sum_{{\bf r}_n\in\Lambda} 
e^{i({\bf q}_\parallel - {\bf k}_\parallel)\cdot{\bf r}_{\parallel n}}
\right]\, 
d{\bf q}_\parallel
\nonumber\\
&=&
- \frac{i}{2 v_0} 
\sum_{{\bf k}_n\in\Lambda^*} 
\frac{e^{i ({\bf k}_\parallel+{\bf k}_n)\cdot {\bf R}_\parallel
+ i K_{\perp n}|R_\perp|} }{K_{\perp n}}  
\nonumber\\
&=& \frac{|R_\perp|}{2 v_0} 
\sum_{{\bf k}_n\in\Lambda^*} 
e^{i ({\bf k}_\parallel+{\bf k}_n)\cdot {\bf R}_\parallel}\,
h_0^{(1)} (K_{\perp n}|R_\perp|).
\label{32dual}
\end{eqnarray}
where $K_{\perp n}$ is given by Eq. (\ref{Kkn}).
Here in going from the first to second equality 
the Poisson  formula (\ref{poisd}) 
has been applied.
Note in passing that the respective dual representations 
for a 1D periodicity in 2D
and 2D periodicity in 3D are formally identical, the only difference being
the dimensionality of $\Lambda^*$ in (\ref{32dual}).

Since exponentials in the second equality in (\ref{32dual}) can be rewritten as
a product of two ``eigenfunctions",
the dual representation can be recast in the form of
an eigenfunction expansion of $G_{0\Lambda}$ [cf. Eq. (\ref{dualbulk})]. 
On the other hand, since $H_0(K_{\perp n} |R_\perp|)$ in Eq. (\ref{31dual})
cannot be factored out as a product of two ``eigenfuctions",
an eigenfunction interpretation of 
a dual representation is obscured for the case of a 1D periodicity in 3D.

\subsection{Convergence and limiting cases}
Note that, according to Eq. (\ref{Kkn}), $K_{\perp n}$ is purely imaginary
with positive imaginary part for $|{\bf k}_\parallel+{\bf k}_n| >\sigma $.
For purely imaginary argument $z=ix$ with $x>0$ the Hankel functions
$H_{0}^{(1)} (ix)$ are related to modified Bessel functions 
$K_{0}(x)$ (Eq. (9.6.4) of Ref. \cite{AS}), whereas $h_{0}^{(1)} (ix)$
in the series  for $d-d_\Lambda=2$ are related 
to modified Bessel functions of third kind
 $\sqrt{\pi/(2x)} K_{1/2}(x)$ (Eq. (10.2.15) of Ref. 
\cite{AS}). This results in rapidly decaying terms and exponential convergence
(see Eqs. (9.7.2) and (10.2.17) of Ref. \cite{AS}).
Exponential convergence can also be directly inferred from
the explicit expression for
\begin{equation}
 h_0^{(1)}(z)=\frac{e^{iz}}{iz}\cdot
\label{h0z}
\end{equation}
One then easily finds that
the respective terms in series
(\ref{32dual}) become exponentially decreasing with
increasing $|{\bf k}_\parallel+{\bf k}_n|$ for $R_\perp\neq 0$.
Since $H_{0}^{(1)}(z) \sim \sqrt{2/(\pi z)}\, e^{i(z-\pi/4)}$ for $|z|\rightarrow\infty$
and $-\pi< \mbox{arg}\, z<2\pi$
(see Eq. (9.2.3) of Ref. \cite{AS}; see also Eq. (\ref{hlas}) above), 
similar applies to the series (\ref{31dual}).
Therefore, although the  convergence  of a dual representation of Green's function
is initially (for $|{\bf k}_\parallel+{\bf k}_n| \leq \sigma $) slow as the 
series consists of mere oscillating terms,
afterward (for $|{\bf k}_\parallel+{\bf k}_n|>\sigma$)  
convergence becomes exponential for $R_\perp\neq 0$ (assuming as usual 
$K_{\perp n}\neq 0$).

Note in passing that all the dual representations of the reduced sums 
are {\em nonanalytic} in $R_\perp$ as they are functions of $|R_\perp|$.
In the case of series (\ref{32dual}) one has
\begin{equation}
|R_\perp| h_{0}^{(1)}(K_{\perp n}|R_\perp|) \rightarrow -i/K_{\perp n}
\hspace*{1.5cm} \mbox{as }|R_\perp| \rightarrow 0.
\nonumber
\end{equation}
Therefore, absolute convergence
of the  resulting series in the limiting case cannot be established 
even for a 1D periodicity in 2D (\cite{WWat}, pp. 51-52). 
Even worse, in the case of a  1D periodicity in 3D individual terms of the
series (\ref{31dual}) possess a logarithmic singularity 
(see Eqs. (9.1.3), (9.1.13) of Ref. \cite{AS}).

To this end, it has been demonstrated that absolute convergence
of dual representations can only be established under the assumption of
$R_\perp\neq 0$. Additionally, obtaining 
the quasi-periodic Green's function of the Laplace 
equation from that of the Helmholtz equation  by taking the limit 
$\sigma \rightarrow 0$ in the resulting
expressions (see Sec. \ref{sec:pois} below) is problematic 
when starting from a dual representation. 
Since
\begin{equation}
{\cal H}_{0}^{+}(K_{\perp n}|R_\perp|) 
     \rightarrow{\cal H}_{0}^{+} (i |{\bf k}_\parallel+{\bf k}_n| |R_\perp|)
\hspace*{1.5cm} \mbox{as }\sigma \rightarrow 0,
\label{lplim}
\end{equation}
absolute convergence can again be established 
only for $R_\perp\neq 0$.
In order to resolve the above problems
 it turns out expedient  to 
invoke representations which converge uniformly and absolutely with respect
to ${\bf R}$. Such representations are known
as the Ewald representations \cite{Ew1,Ew2}.
Additional bonus of the Ewald  representations is that they enable one to
investigate analytic properties of quasi-periodic Green's functions
in the complex variable $z=\sigma^2$.

As a final remark of this section note that dual 
representations in the quasi-periodic case
can also be established by applying 
an Ewald integral representation of Green's function
and the generalized Jacobi identity.
This path, which has been originally followed by McRae 
\cite{McRaer} for a 2D periodicity in 3D,
is outlined in Appendix \ref{app:fsqpgf}.

\section{Ewald representations of quasi-periodic $G_{0\Lambda}$}
\label{sec:ew} 
In going from the spatial domain form [Eq. (\ref{lgrfdf})] to 
the respective spectral domain forms of $G_{0\Lambda}$
[Eqs. (\ref{31dual}), (\ref{32dual})],
the summation over $\Lambda$ has been fully replaced by a summation over $\Lambda^*$.
In this section, starting from the
 spectral domain forms of $G_{0\Lambda}$ a half-step backward  will be performed 
resulting in a hybrid Ewald
representation of $G_{0\Lambda}$.  The Ewald
representation of $G_{0\Lambda}$ involves sums over both 
$\Lambda$ and $\Lambda^*$ and, in contrast to a dual, spectral domain, 
form of $G_{0\Lambda}$,  
is valid for all $R_\perp$ and uniformly
convergent with respect to bounded sets of ${\bf R}$, 
provided that ${\bf R}\not\in \Lambda$.

In deriving the Ewald
representation of $G_{0\Lambda}$, we first recall 
 formulae (10.1.1) and (9.1.6) of Ref. \cite{AS} and recast $h_0^{(1)}$
in the series (\ref{32dual}) as
\begin{equation}
h_0^{(1)}(z) =\left(
\frac{\pi}{2z}
\right)^{1/2} H_{1/2}^{(1)}(z)=-i\left(
\frac{\pi}{2z}
\right)^{1/2} H_{-1/2}^{(1)}(z).
\end{equation}
Therefore, a dual representation of $G_{0\Lambda}$ 
in any quasi-periodic case [Eqs. (\ref{31dual}), (\ref{32dual})]
can be rewritten as a sum of  
{\em cylindrical} Hankel functions of an appropriate order.

Now for $d-d_\Lambda=2$ we shall introduce 
\begin{equation} 
G_{\nu \Lambda}(\sigma,{\bf k},{\bf R})= -\frac{i}{4v_0} \sum_{{\bf k}_n\in\Lambda^*}
 \left(\frac{|R_\perp|}{K_{\perp n}}\right)^{\nu/2} \, 
H_{-\nu/2}^{(1)}(K_{\perp n} |R_\perp|) 
 e^{i({\bf k}_\parallel+{\bf k}_n)\cdot{\bf R}_\parallel}.
\label{anf1in3}
\end{equation}
In the following, $G_{\nu \Lambda}$ will be called an {\em analytic form} 
of $G_{0\Lambda}$. Obviously $G_{0\Lambda}= \left. G_{\nu \Lambda}\right|_{\nu=0}$.
Next it turns out expedient to employ
the following integral representation [see Eq. (\ref{kmb32})
of Appendix \ref{app:hintr}],
\begin{equation} 
\left(\frac{|R_\perp|}{K_{\perp n}}\right)^{\nu/2} \, 
H_{-\nu/2}^{(1)}(K_{\perp n} |R_\perp|) 
= \frac{1}{\pi i} \int_{0_+}^{\infty 
\exp{i\phi_n}} \zeta^{\nu/2-1} e^{\frac{1}{2}\, 
(K_{\perp n}^2 \zeta - |R_\perp|^2/\zeta)}\, d\zeta.
\label{hs32} 
\end{equation}  
The lower limit $0_+$ indicates that the contour integral starts from $0$
in the direction of the positive real axis. 
Here we have used the convention (see Ref. \cite{WWat}, p. 589) that
\begin{equation}
\left(\frac{|R_\perp|}{K_{\perp n}}\right)^{\nu/2} =
\exp\left\{
\frac{\nu}{2}\left(
\ln\left|\frac{R_\perp}{K_{\perp n}}\right|
   -i\arg K_{\perp n}\right)
\right\},
\end{equation}
where the argument of $K_{\perp n}$ takes the values $0$ or $\pi/2$, and
 $\phi_n$ is given by 
\begin{equation}
\phi_n =\pi -2 \arg K_{\perp n}.
\label{phin}
\end{equation}

From Eq. (\ref{hs32}) it follows that 
$H_{0}^{(1)}(K_{\perp n} |R_\perp|)$, and hence also $G_{0\Lambda}$, can 
be analytically continued in the complex parameter $\nu$.
This explains the reason why $G_{\nu \Lambda}$ defined by Eq. (\ref{anf1in3})
has been called an {\em analytic form} of $G_{0\Lambda}$.
Indeed, as soon as Re $\nu>0$,
\begin{equation}
H_{-\nu/2}^{(1)} (z) \sim  - i\frac{e^{\nu\pi i/2}}{\pi} \Gamma(\nu/2) (z/2)^{-\nu/2}
  \hspace*{1.5cm} \mbox{as }z\rightarrow 0
\label{hnuas}
\end{equation}
(Eqs. (9.1.6) and  (9.1.9) of Ref. \cite{AS}).
Therefore, all the terms in the series (\ref{anf1in3}) for Re $\nu>0$
are singularity free as $|R_\perp|\rightarrow 0$,
 and the original 
logarithmic singularity of $H_{0}^{(1)}(K_{\perp n}|R_\perp|)$ in the limit
 $|R_\perp|\rightarrow 0$ in the dual representation  (\ref{31dual}) of $G_{0\Lambda}$ 
is thereby avoided. 
Additionally, the series in (\ref{anf1in3}) can be easily seen as an analytic function
of $\nu$ for all values of ${\bf R}$, provided that Re $\nu> 2d_\Lambda$. 
In the latter case, upon using asymptotic (\ref{hnuas}), 
the elementary products $|R_\perp|^{\nu/2} 
H_{-\nu/2}^{(1)}(K_{\perp n} |R_\perp|)$ in the series (\ref{anf1in3})
can be uniformly bounded for all $n$ by a {\em finite} number as $|R_\perp|\rightarrow 0$.
On the other hand, the asymptotics of 
$H_{-\nu/2}^{(1)}(z)= e^{\nu\pi i/2}H_{\nu/2}^{(1)}(z)$ (Eq. (9.1.6) of Ref. \cite{AS})
as $K_{\perp n} \rightarrow \infty$ is determined according to Eq. (\ref{hlas}) with $d=2$.
Consequently, the series (\ref{anf1in3}) can be 
uniformly bounded by the series $K_{\perp n}^{-\nu/2}$ for all ${\bf R}$.
Now the series $K_{\perp n}^{-\nu/2}$ 
is absolutely convergent for Re $\nu> 2d_\Lambda$ (Ref. \cite{WWat}, pp. 51-52).
Therefore, since the sum in  Eq. (\ref{anf1in3}) 
is absolutely and uniformly convergent for all ${\bf R}$, it
defines an analytic function of $\nu$ for all values 
of $R_\perp$ and ${\bf R}_\parallel$ if Re $\nu > 2d_\Lambda$ 
(Ref. \cite{WWat}, Sec. 5.32).

Now, the task is to find an analytic continuation $G_{\nu \Lambda}$ 
from a domain Re $\nu > 2d_\Lambda$ to a domain containing $\nu=0$.
Here it has been implicitly assumed (and will be proved later on)
that the analytical 
of $G_{\nu \Lambda}$ defined for Re $\nu > 2d_\Lambda$ by the series
(\ref{anf1in3}) will yield our $G_{0 \Lambda}$ defined
by Eq. (\ref{lgrfdf}). 
The necessity of
an analytic  continuation in the quasi-periodic case 
makes derivation of the Ewald representation of $G_{0\Lambda}$
fundamentally different from that in the bulk case.
This analytical continuation procedure is analogous to finding 
the values of the Riemann $\zeta$-function outside the domain 
of absolute convergence, Re $\nu\leq 1$, of its defining series
\begin{equation}
\zeta(\nu)=\sum_{n=1}^\infty n^{-\nu}.
\end{equation}
In order to find out an analytic continuation $G_{\nu \Lambda}$ 
in a domain containing $\nu=0$, one substitutes (\ref{hs32}) 
back into (\ref{anf1in3}) which results in
\begin{eqnarray}
G_{\nu \Lambda}(\sigma,{\bf k}_\parallel,{\bf R}) &=&
-\frac{1}{4\pi v_0} 
\sum_{{\bf k}_n\in\Lambda^*}^{K_{\perp n}^2>0} 
    e^{i({\bf k}_\parallel+{\bf k}_n)\cdot{\bf R}_\parallel}
  \int_{0_+}^{-\infty }
\zeta^{\nu/2-1} e^{\frac{1}{2}\, 
       (K_{\perp n}^2 \zeta- |R_\perp|^2/\zeta)}\,
d\zeta\nonumber\\
&&
-\frac{1}{4\pi v_0} 
\sum_{{\bf k}_n\in\Lambda^*}^{K_{\perp n}^2<0} 
     e^{i({\bf k}_\parallel+{\bf k}_n)\cdot{\bf R}_\parallel}
  \int_{0_+}^{+\infty}
\zeta^{\nu/2-1} 
   e^{\frac{1}{2}\, (K_{\perp n}^2 \zeta- |R_\perp|^2/\zeta)}\,
d\zeta. 
\nonumber\\
 \label{31grfss}
\end{eqnarray}
The first sum has only a limited number of terms so that the order
of integration and summation can be inverted.
By a straightforward generalization of Riemann's method
(see Ref. \cite{WWat}, p. 273),
the inversion of integration and summation
also holds for the second sum in
(\ref{31grfss}), provided that Re $\nu> 2d_\Lambda$ and if always 
Re $\zeta>0$ on the contour of integration \cite{Kam1}.

The remaining  two steps in the derivation of the 
Ewald representation
are essentially those used by Epstein in 
an analytic continuation of his 
zeta functions \cite{Ep,GlZ}:

\begin{itemize}

\item 
the resulting contour integral is split in two parts 
by taking a point $\eta$ somewhere in the domain
Re $\eta>0$, $|\eta|<\infty$. 

\item 
 the 
generalized Jacobi identity (\ref{edi}), which is valid for
Re $\zeta>0$,  is applied 
for the part of the integral from $0$ to $\eta$ with $\zeta=1/(2\xi^2)$,
and ${\bf r}_s=-{\bf r}_s$,  yielding
\begin{equation}
\sum_{{\bf k}_n\in\Lambda^*}
e^{-({\bf k}_n+{\bf k}_\parallel)^2 \zeta/2 
         +i({\bf k}_\parallel+{\bf k}_n)\cdot{\bf R}_\parallel} 
=
\frac{v_0}{(2\pi\zeta)^{d_\Lambda /2}}
\sum_{{\bf r}_n\in\Lambda} 
e^{-({\bf R}_\parallel+{\bf r}_n)^2/(2\zeta) 
                         - i{\bf k}_\parallel\cdot{\bf r}_n}
\label{edikmb}
\end{equation}

\end{itemize}

Following the first step, $G_{\nu \Lambda}$ is expressed as the sum of two terms,
\begin{equation}
G_{\nu\Lambda}(\sigma,{\bf k}_\parallel,{\bf R})
= G_1(\sigma,{\bf k}_\parallel,{\bf R})
 + G_2(\sigma,{\bf k}_\parallel,{\bf R}),
\label{gisg1p2}\
\end{equation}
where the respective $G_1$ and $G_2$ contributions 
result from the respective contour integrals over $(0,\eta)$ and $(\eta,\infty)$. 
Obviously, although each of the partial integrals depends on $\eta$, 
called the Ewald  parameter, their sum does not.

Following the second step, Eq. (\ref{31grfss}) is transformed into
\begin{eqnarray}
G_{\nu \Lambda}({\bf R}) &=&
-\frac{1}{4\pi v_0} 
\int_{\eta}^{-\infty } \sum_{{\bf k}_n\in\Lambda^*}^{K_{\perp n}^2>0} 
    e^{i({\bf k}_\parallel+{\bf k}_n)\cdot{\bf R}_\parallel}
   \zeta^{\nu/2-1} 
e^{\frac{1}{2}\, (K_{\perp n}^2 \zeta- |R_\perp|^2/\zeta)}\,
d\zeta\nonumber\\
&&
-\frac{1}{4\pi v_0} 
\int_{\eta}^{+\infty} \sum_{{\bf k}_n\in\Lambda^*}^{K_{\perp n}^2<0} 
   e^{i({\bf k}_\parallel+{\bf k}_n)\cdot{\bf R}_\parallel}
   \zeta^{\nu/2-1} 
e^{\frac{1}{2}\, (K_{\perp n}^2 \zeta- |R_\perp|^2/\zeta)}\,
d\zeta\nonumber\\
&&
-\frac{1}{(2\pi)^{3/2} } \left(\frac{\pi}{2} \right)^{1/2}
 \int_{0}^{\eta}
       \sum_{{\bf r}_n\in\Lambda} e^{-i{\bf k}_\parallel\cdot{\bf r}_n}
\zeta^{\nu/2-3/2} e^{\frac{1}{2}\, \left[ \sigma^2 \zeta - 
                        ({\bf R}+{\bf r}_n)^2/\zeta\right]}\,
d\zeta.
\end{eqnarray}
The latter expression is an analytic function of $\nu$ for all values 
of $\nu$ if $|{\bf R}|\neq 0$, or more generally
if ${\bf R}\not\in \Lambda$, and it represents the sought 
analytic continuation of (\ref{anf1in3}) 
for Re $\nu\leq 2 d_\Lambda$. Note in passing that for ${\bf R}\in \Lambda$ analyticity 
can only be established if Re $\nu > 1$. Otherwise, the last integral diverges
for ${\bf r}_n=-{\bf R}$.

On putting $\nu=0$, and hence assuming ${\bf R}\not\in \Lambda$, inverting again 
the order of summation and integration (since it can be allowed), and, substituting 
$\zeta\rightarrow 1/\zeta$ in the last integral,
\begin{eqnarray}
G_{0\Lambda}(\sigma,{\bf k}_\parallel,{\bf R}) &=& 
- \frac{1}{4\pi v_0} 
\sum_{{\bf k}_n\in\Lambda^*} 
e^{i({\bf k}_\parallel+{\bf k}_n)\cdot{\bf R}_\parallel}
  \int_{\eta}^{\infty \exp{i\phi_n}}
\zeta^{-1} 
e^{\frac{1}{2}\, (K_{\perp n}^2 \zeta- |R_\perp|^2/\zeta)}\,
d\zeta\nonumber\\
&&
-\frac{1}{4\pi^2} \left(\frac{\pi}{2} \right)^{1/2}
 \sum_{{\bf r}_s\in\Lambda} e^{-i{\bf k}_\parallel \cdot{\bf r}_s}
  \int_{1/\eta}^{\infty}
\zeta^{-1/2} e^{\frac{1}{2}\, \left[\sigma^2 /\zeta - 
                         ({\bf R}+{\bf r}_s)^2\zeta\right] }\,
d\zeta.
\label{31grfr}
\end{eqnarray}
The restriction Re $\zeta>0$ going back to the integral
representation (\ref{hs32}) can be now removed.

\subsection{Complementary cases}
For completeness, for $d-d_\Lambda=1$ one would begin with 
the analytic form 
\begin{equation}
G_{\nu \Lambda}(\sigma,{\bf k}_\parallel,{\bf R}) = -\frac{i}{2v_0} 
\left(\frac{\pi}{2} \right)^{1/2}
\sum_{{\bf k}_n\in\Lambda^*} 
\left(\frac{|R_\perp|}{K_{\perp n}}\right)^{\nu/2} \,
H_{-\nu/2}^{(1)}(K_{\perp n} |R_\perp|) \,
 e^{i({\bf k}_\parallel+{\bf k}_n)\cdot{\bf R}_\parallel}.
\label{32grfs}
\end{equation}
Similarly as in the preceding case, 
$G_{\nu \Lambda}$ defines for Re $\nu > 2d_\Lambda$ an analytic function 
of $\nu$ for all values of $R_\perp$ and ${\bf R}_\parallel$, provided that it 
remains ${\bf R}\not\in \Lambda$.
The task is now to find an analytic continuation $G_{\nu \Lambda}$ 
from a domain Re $\nu> 2d_\Lambda$ to a domain containing $\nu=1$.
After repeating the steps which led from Eq. (\ref{anf1in3}) to Eq.
(\ref{31grfr}), on putting $\nu=1$ and inverting again the order 
of summation and integration, one finds 
for $d_\Lambda=2$, $d=3$ \cite{Kam3,Kam2}
\begin{eqnarray}
G_{0\Lambda}(\sigma,{\bf k}_\parallel,{\bf R}) &=&
-\frac{1}{2\pi v_0} \left(\frac{\pi}{2} \right)^{1/2}
\sum_{{\bf k}_n\in\Lambda^*} 
   e^{i({\bf k}_\parallel+{\bf k}_n)\cdot{\bf R}_\parallel}
  \int_{\eta}^{\infty \exp{i\phi_n}}  \zeta^{-1/2} 
e^{\frac{1}{2}\, (K_{\perp n}^2 \zeta- |R_\perp|^2/\zeta)}\,
d\zeta\nonumber\\
&&
-\frac{1}{4\pi^2} \left(\frac{\pi}{2} \right)^{1/2}
     \sum_{{\bf r}_s\in\Lambda} 
      e^{-i{\bf k}_\parallel \cdot{\bf r}_s}
  \int_{1/\eta}^{\infty}
\zeta^{-1/2} e^{\frac{1}{2}\, \left[ \sigma^2 /\zeta - 
                           ({\bf R}+{\bf r}_s)^2\zeta\right] }\,
d\zeta.
\label{32grfr}
\end{eqnarray}
Similarly, for $d_\Lambda=1$, $d=2$ one arrives (see, e.g., Ref. \cite{Lin}) at 
\begin{eqnarray}
G_{0\Lambda}(\sigma,{\bf k}_\parallel,{\bf R}) 
&=& - \frac{1}{2\pi v_0} \left(\frac{\pi}{2} \right)^{1/2}
\sum_{{\bf k}_n\in\Lambda^*} 
   e^{i({\bf k}_\parallel+{\bf k}_n)\cdot{\bf R}_\parallel}
  \int_{\eta}^{\infty \exp{i\phi_n}}
\zeta^{-1/2} 
 e^{\frac{1}{2}\, (K_{\perp n}^2 \zeta- |R_\perp|^2/\zeta)}\,
d\zeta\nonumber\\
&&
- \frac{1}{4\pi} \sum_{{\bf r}_s\in\Lambda} 
    e^{-i{\bf k}_\parallel \cdot{\bf r}_s}
  \int_{1/\eta}^{\infty}
\zeta^{-1} e^{\frac{1}{2}\, \left[ \sigma^2 /\zeta - 
                           ({\bf R}+{\bf r}_s)^2\zeta\right] }\,
d\zeta.
\label{21grfr}
\end{eqnarray}

It can be proved directly (see, e.g., Appendix 3 of Ref. \cite{Kam1})
that Ewald representations (\ref{31grfr}), (\ref{32grfr}), (\ref{21grfr})
satisfy Eq. (\ref{helmeq}) and the boundary conditions. Therefore,
they are required Green's function, which provides 
a posteriori justification of the outlined analytic 
continuation procedure.

Since the respective dual representations for a 1D periodicity in 2D
and 2D periodicity in 3D are formally identical, the terms involving 
a sum over reciprocal lattice in
Eqs.  (\ref{32grfr})  and (\ref{21grfr}) are identical.
Surprisingly enough,  the terms involving a sum over direct lattice in
Eqs. (\ref{31grfr}) and (\ref{32grfr}) are identical too.

\subsection{Ewald vs dual representations}
\begin{figure}[tbp]
\begin{center}
\epsfig{file=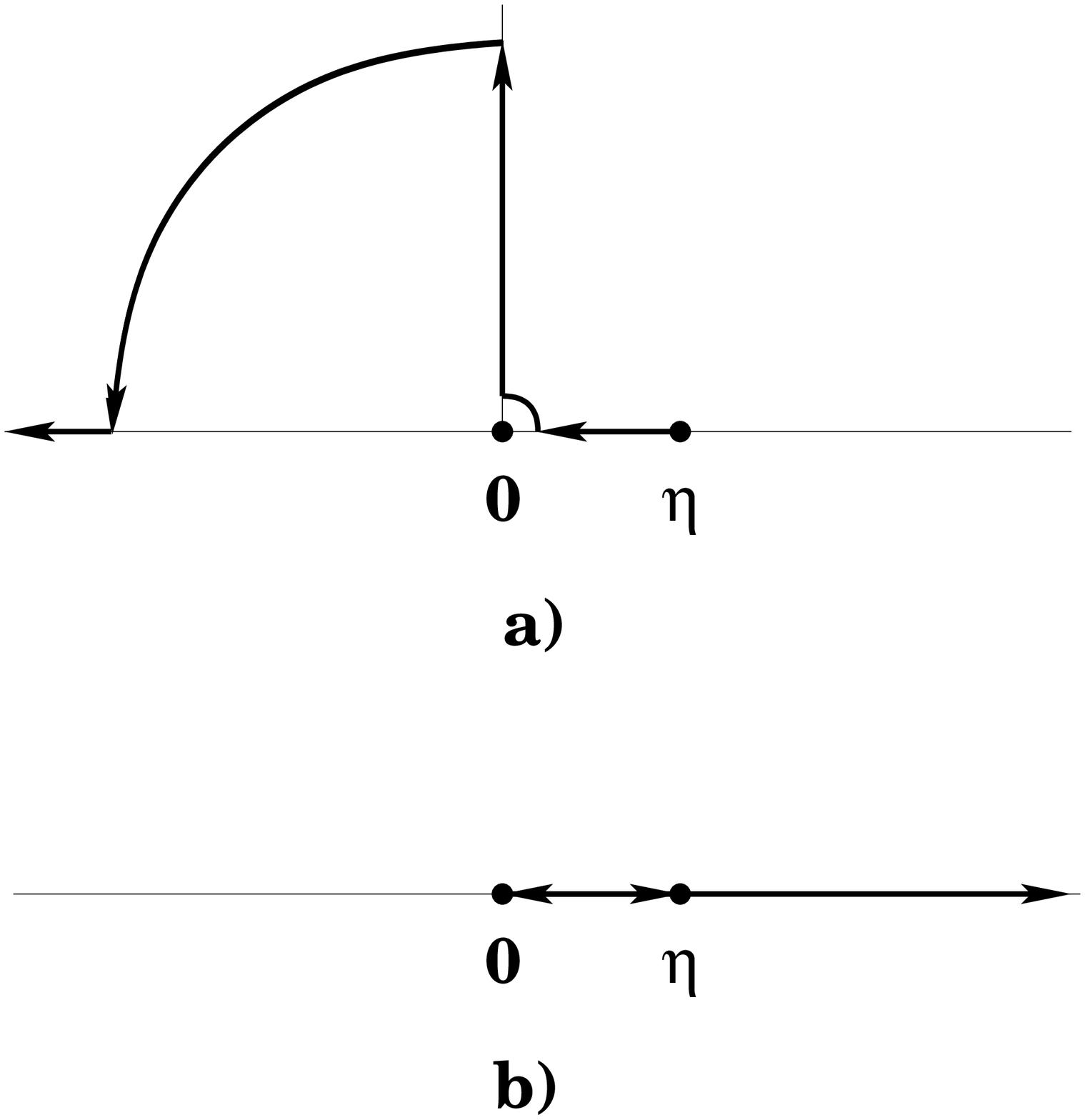,width=12cm,clip=0,angle=0}
\end{center}
\caption{}
\label{fg:kmb}
\end{figure}

One has (see Appendix 1 of Ref. \cite{Kam1})
\begin{eqnarray}
\lefteqn{
\int_{\eta}^{\infty \exp{i\phi_n}}
\zeta^{-1/2} 
 e^{\frac{1}{2}\, (K_{\perp n}^2 \zeta- |R_\perp|^2/\zeta)}\,
d\zeta =
}\nonumber\\
&& 
-\sqrt{2\pi}\, \frac{e^{i  K_{\perp n}|R_\perp|} }{iK_{\perp n}}
- 
\int_{1/\eta}^\infty 
\zeta^{-3/2} 
 e^{\frac{1}{2}\, (K_{\perp n}^2/ \zeta- |R_\perp|^2 \zeta)}\,
d\zeta
\nonumber
\end{eqnarray}
and
\begin{eqnarray}
\lefteqn{
\int_{\eta}^{\infty \exp{i\phi_n}}
\zeta^{-1} 
 e^{\frac{1}{2}\, (K_{\perp n}^2 \zeta- |R_\perp|^2/\zeta)}\,
d\zeta =
}\nonumber\\
&& 
\pi i H_0^{(1)} ( K_{\perp n}|R_\perp| )
- 
\int_{1/\eta}^\infty 
\zeta^{-1} 
 e^{\frac{1}{2}\, (K_{\perp n}^2/ \zeta- |R_\perp|^2 \zeta)}\,
d\zeta.
\nonumber
\end{eqnarray}
For $K_{\perp n}^2>0$ this can be shown by deforming integration contour 
in Eqs. (\ref{31grfr}), (\ref{32grfr}), (\ref{21grfr}) to that shown
in Fig. \ref{fg:kmb}{\bf a} and upon
invoking Jordan's lemma for the integration along quarter circles 
(\cite{WWat}, p. 115). For $K_{\perp n}^2<0$ one 
then takes the contour as shown
in Fig. \ref{fg:kmb}{\bf b}.
Therefore, a  comparison of dual representations (\ref{31dual}), (\ref{32dual}) 
of quasi-periodic free-space Green's function with their respective
Ewald representations (\ref{31grfr}), (\ref{32grfr}), (\ref{21grfr})
 shows that to each term of a dual representation 
the second term (integral above) is added
to make the series convergent uniformly with respect to $R_\perp$
(provided that ${\bf R}\not\in \Lambda$).
These terms are then compensated by the series over $\Lambda$.

For a sufficiently 
large $\eta$, $G_{0\Lambda}$ can often be well approximated by the 
series over $\Lambda$. 
This approximation to $G_{0\Lambda}$
is called the {\em incomplete} Ewald summation \cite{KR}.

\section{Calculation of the lattice sums $D_{L}$} 
\label{sec:dlcalc}  
The lattice sums $D_{L}$ have been defined by Eq. (\ref{dstrco})
as the expansion coefficients of $G_{0\Lambda}$ in terms of the  
regular (cylindrical in 2D, spherical in 3D) waves, 
or, alternatively, as the 
Schl\"{o}milch series (\ref{dstrcos}).
Analytic closed expression of the lattice sums $D_{L}$
can only be obtained 
in the particular case of $l=0$ and 
a 1D lattice $\Lambda$ with a period $a$ 
in 3D \cite{Gla3,GHM}.
Indeed, assuming the elementary identity 
\begin{equation}
\ln(1-z) = - \sum_{n>0}\frac{z^n}{n},
\nonumber
\end{equation}
one obtains 
\begin{eqnarray}
\lefteqn{
 \sum_{{\bf r}_n\in\Lambda}{}^{'}   e^{i{\bf k}_\parallel\cdot{\bf r}_n}
\frac{e^{i\sigma r_n} }{r_n} 
= \frac{1}{a} \sum_{n\neq 0}{}^{'}  e^{ia {\bf k}_\parallel n }
\frac{e^{i\sigma a |n|} }{|n|}
}
\nonumber\\
&&
=- \frac{1}{a}\left\{
\ln\left[ 1- e^{ia (\sigma- {\bf k}_\parallel) }\right]
+
\ln \left[ 1- e^{ia (\sigma+ {\bf k}_\parallel) }\right]
\right\}
\nonumber\\
&&
=- \frac{1}{ a}
\ln\left[e^{2ia\sigma} - 2\cos({\bf k}_\parallel a) e^{ia \sigma} + 1 \right].
\label{exct1in3}
\end{eqnarray}
Hence, upon using that $-iACY_{00}=-i\sigma/\sqrt{4\pi}$ in 3D,
\begin{eqnarray} 
 D_{00}(\sigma,{\bf k}_\parallel) &=& -i \frac{\sigma}{\sqrt{4\pi} }  
+ \frac{1}{\sqrt{4\pi} a}
\ln\left[e^{2ia\sigma} - 2\cos({\bf k}_\parallel a) e^{ia \sigma} + 1 \right]
\nonumber\\
&=&    \frac{1}{\sqrt{4\pi} a}
\ln\left\{ 2\left[\cos(\sigma a) - \cos({\bf k}_\parallel a)  \right]\right\},
\label{dstrcos0} 
\end{eqnarray}
which is up to the prefactor of $-1/(\sqrt{4\pi} a)$ 
the $\hat{\gamma}$-function of Karpeshina (see Eq. (29) of
Ref. \cite{Kar} for $\sigma=is$). 
(When solving Eq. (\ref{zps}), the principal branch 
of logarithm is assumed
in the above expression for $D_{00}$ and our spectral parameter $\tilde{\alpha}$
has also been rescaled compared to that of
Karpeshina \cite{Kar,Krp} with the above prefactor.)
In accordance with our notation, boldface ${\bf r}_n$ and ${\bf k}_\parallel$
are numbers which can be either positive or negative, whereas
$r_n\geq 0$ stands for absolute value.
(It is reminded here that the energy operator 
in the one-particle theory of periodic point interactions 
is constructed in terms of the operator of multiplication by 
$D_{00}$ ($\gamma$ function of Karpeshina \cite{Kar,Krp}) and that
$D_{00}$ determines the spectrum according to Eq. (\ref{zps}).)

Invoking that 
$D_{L}{}'s$ are independent of $R$, the lattice sums in all remaining cases 
are calculated as \cite{Kam3,Kam2,Mol}   
\begin{equation}
D_{L}(\sigma,{\bf k}_\parallel)
 = \lim_{R\rightarrow 0}   \frac{1}{ {\cal J}_{|l|}(\sigma R)} 
\oint  {\cal Y}_L^*(\hat{{\bf R}})    
D_\Lambda(\sigma,{\bf k}_\parallel,{\bf R}) \,d\Omega_{\bf R},   
\label{dlcalc}   
\end{equation}    
where $\oint d\Omega_{\bf R}$ denotes the angular 
integration  over all directions of ${\bf R}$.  
In calculating $D_{L}$, the respective
Ewald representations (\ref{31grfr}), (\ref{32grfr}), (\ref{21grfr}) of 
 $G_{0\Lambda}$ are substituted in the defining Eq. (\ref{dlmbd}) 
for $D_\Lambda(\sigma,{\bf k}_\parallel,{\bf R})$.
The two series in the respective
Ewald representations (\ref{31grfr}), (\ref{32grfr}), (\ref{21grfr})
are uniformly convergent with respect to ${\bf R}$ so that
the series can be term-wise integrated when $D_{L}$ is calculated according
to Eq. (\ref{dlcalc}).
Following a hybrid character of the 
Ewald representations (\ref{31grfr}), (\ref{32grfr}), (\ref{21grfr}),
the respective $D_{L}$ are conventionally written 
as a sum \cite{HS,Kam3,Kam2,Mol,Moc} 
\begin{equation}  
D_L(\sigma,{\bf k}_\parallel) = 
  D_L^{(1)}(\sigma,{\bf k}_\parallel)+ 
  D_L^{(2)}(\sigma,{\bf k}_\parallel)+   
  D_L^{(3)}(\sigma),   
\label{dldsum}  
\end{equation} 
 where $D_{L}^{(1)}$ ($D_{L}^{(2)}$) involves a sum over  
 reciprocal lattice (all ${\bf r}_n\neq 0$ terms of the direct lattice).  
$D_{L}^{(3)}$ is the term which  combines $G_0^p({\bf R})$ and the 
 ${\bf r}_n=0$ contribution of the direct lattice sum $G_2$.   
$D_{L}^{(3)}$ is only nonzero for $l=0$,  
\begin{equation}  
D_{L}^{(3)}=D_{0}^{(3)}\delta_{L0}.  
\label{dl3ex}  
\end{equation}   

In the following, the respective contributions $D_{L}^{(1)}$, $D_{L}^{(2)}$, and
$D_{L}^{(3)}$ will be calculated. For reader not interested in 
an explicit derivation of results,
the resulting expressions are given by 
Eqs. (\ref{dl1f1in3}), (\ref{dl2f12in3}), and (\ref{dl3f1in3})
[see Eqs. (\ref{dl1f2in3}), (\ref{dl2f12in3}), (\ref{dl3f1in3}) 
for $d_\Lambda=2$, $d=3$ and Eqs.  (\ref{dl1f1in2}),  (\ref{dl2f1in2}), 
(\ref{dl3f1in2}) for $d_\Lambda=1$, $d=2$].

\subsection{Consequences of the reflection symmetry 
for the lattice sums in  the quasi-periodic case}  
\label{sec:reflsym}
Assuming standard spherical coordinates, one has 
\begin{equation} 
Y_L(\hat{{\bf R}}_\parallel-\hat{{\bf R}}_\perp)= 
(-1)^{l+m} Y_{L}(\hat{{\bf R}}_\parallel+\hat{{\bf R}}_\perp).
\label{3dshrref} 
\end{equation}
Therefore, for    
the lattice plane perpendicular to the $z$-axis
\begin{equation} 
D_{L}\equiv 0, \hspace*{2cm} l+m\hspace*{0.1cm}\mbox{odd} 
\label{3ddgenpr} 
\end{equation}  
for both $d_\Lambda=1$, $d=3$ and $d_\Lambda=2$, $d=3$ cases.   
This identity follows upon combining the property (\ref{ordpr}) 
with the expansion (\ref{dstrco}). In fact [see Eq. (\ref{dl3ex})
and Eqs. (\ref{3dd1genpr}), (\ref{3dd2genpr}) below],
it will be shown that (for the  lattice plane perpendicular
to the $z$-axis) the property (\ref{3ddgenpr}) holds for each of
the contributions 
$D_L^{(j)}$, $j=1,2,3$,  separately.

\subsection{Calculation of $D_{L}^{(1)}$} 
\label{sec:lyardl1} 

\subsubsection{General part} 
As it has been alluded to above, the contribution $D_{L}^{(1)}$ 
derives from the sum over reciprocal lattice $\Lambda$ in 
the corresponding Ewald representation of $G_{0\Lambda}$.
According to the Ewald representations 
(\ref{31grfr}), (\ref{32grfr}), (\ref{21grfr}) of the 
quasi-periodic Green's functions, 
\begin{equation}
D_{L}^{(1)} = - \frac{1}{2 v_0(2\pi)^c}
I_L^{(1)},
\label{d1fcod1}
\end{equation}
where
\begin{equation}
I_L^{(1)}= 
\lim_{R\rightarrow 0} \frac{1}{j_{|l|}(\sigma R)} 
\oint  Y_L^*(\hat{{\bf R}}) 
   e^{i({\bf k}_\parallel+{\bf k}_s)\cdot{\bf R}_\parallel}
\int_{\eta}^{\infty \exp{i\phi_n}}
\zeta^{-c} e^{\frac{1}{2}\, (K_{\perp s}^2 \zeta- |R_\perp|^2/\zeta)}\,
d\zeta \,d\Omega_{\bf R},
\end{equation}
and $1/2\leq c=(d-d_\Lambda) /2\leq 1$. 
The exponential decrease of the integrand with increasing $\zeta$
for the integration over $\zeta$ (assuming as usual $K_{\perp s}\neq 0$) 
guarantees that the order of integration can be inverted.
In order to perform the latter integral, the exponential 
is expanded into a power series of $|R_\perp|^2$
resulting in
\begin{eqnarray}
\lefteqn{
I_L^{(1)} \equiv \sum_{n=0}^\infty \frac{(-1)^n}{2^n n!}
\int_{\eta}^{\infty \exp{i\phi_n}}
\zeta^{-c-n} e^{K_{\perp s}^2 \zeta/2}\, d\zeta}\nonumber\\
&&
\times
\left(\lim_{R\rightarrow 0} \frac{1}{j_{|l|}(\sigma R)} 
\oint  Y_L^*(\hat{{\bf R}}) 
e^{i({\bf k}_\parallel+{\bf k}_s)\cdot{\bf R}_\parallel}
|R_\perp|^{2n}\,d\Omega_{\bf R}\right)\nonumber\\
&&
= \sum_{n=0}^\infty \frac{(-1)^n}{2^n n!}
\left( e^{-\pi i} \frac{K_{\perp s}^2}{2}\right)^{n+c-1} 
\Gamma\left(1-c-n, e^{-\pi i} 
\frac{K_{\perp s}^2 \eta }{2}\right)\, I_\zeta^l(n) \nonumber\\
&&
=2^{1-c} e^{(1-c)\pi i} \, 
\sum_{n=0}^\infty \frac{I_\zeta^l(n) }{2^{2n} n!}\,
 \Gamma\left(1-c-n, e^{-\pi i} \frac{K_{\perp s}^2
\eta }{2}\right) K_{\perp s}^{2(n+c-1)},
\label{il1int}
\end{eqnarray}
where
\begin{equation}
 I_\zeta^l(n) =\lim_{R\rightarrow 0} \frac{1}{j_{|l|}(\sigma R)} 
\oint  Y_L^*(\hat{{\bf R}}) 
  e^{i({\bf k}_\parallel+{\bf k}_s)\cdot{\bf R}_\parallel}
|R_\perp|^{2n}\,d\Omega_{\bf R},
\end{equation}
and $\Gamma$ is the incomplete gamma function (see Eq. (6.5.3) 
of Ref. \cite{AS}).
In the second equality in (\ref{il1int}) we have used in the integral over 
$\zeta$ the substitution
\begin{equation} 
\zeta=\frac{2e^{\pi i}}{K_{\perp s}^2}\, t, 
\end{equation} 
which  leads to
\begin{eqnarray}
\lefteqn{
 \int_{\eta}^{\infty \exp{i\phi_n}} \zeta^{-c-n} 
e^{K_{\perp s}^2 \zeta/2}\, d\zeta =
\left(
-\frac{2}{K_{\perp s}^2}
\right)^{1-c-n}
\int_{e^{-\pi i}K_{\perp s}^2\eta/2}^{\infty} 
t^{-c-n} e^{-t }\, dt
}
\nonumber\\ 
&&
=
\left(
-\frac{2}{K_{\perp s}^2}
\right)^{1-c-n} \,
\Gamma\left(1-c-n, e^{-\pi i} \frac{K_{\perp s}^2 \eta }{2}\right).
\label{incgm}
\end{eqnarray}
In the final equality in (\ref{il1int}) we have substituted
\begin{equation}
\left( e^{-\pi i}
\frac{K_{\perp s}^2}{2}\right)^{n+c-1} = e^{(1-c)\pi i} 2^{1-c}
\frac{(-1)^n}{2^n}\, K_{\perp s}^{2(n+c-1)}.
\end{equation}
Now upon combining Eqs. (\ref{d1fcod1}) and (\ref{il1int})
\begin{eqnarray} 
\lefteqn{ 
D_{L}^{(1)} 
= - \frac{1}{(2\pi)^c} \frac{(-2)^{1-c}}{2 v_0}
\sum_{{\bf k}_s\in\Lambda^*}
\sum_{n=0}^\infty \frac{1}{2^{2n} n!}\,
 \Gamma\left(1-c-n, e^{-\pi i} \frac{K_{\perp s}^2 \eta }{2}\right) 
  K_{\perp s}^{2(n+c-1)}
}
\nonumber\\
&&
\times 
\left(\lim_{R\rightarrow 0} \frac{1}{j_{|l|}(\sigma R)} 
\oint  Y_L^*(\hat{{\bf R}}) 
  e^{i({\bf k}_\parallel+{\bf k}_s)\cdot{\bf R}_\parallel}
|R_\perp|^{2n}\,d\Omega_{\bf R}\right).
\label{dl1prelim}
\end{eqnarray}
It order to finish the calculation of $D_{L}^{(1)}$,
it remains to perform  the angular integration and the 
limit $R\rightarrow 0$. 
In the following this limit will be provided for various
particular cases.

\subsubsection{The case of a 1D periodicity in 3D} 
If the 1D
lattice is oriented along the $x$-axis, ${\bf R}_\perp={\bf R}\cos\theta$,
${\bf R}_\parallel={\bf R}\sin\theta \cos\phi$, and hence
\begin{equation} 
|{\bf R}_\perp| = R |\cos\theta|, \hspace*{1cm} 
({\bf k}_\parallel+{\bf k}_s)\cdot{\bf R}_\parallel= 
|{\bf k}_\parallel+{\bf k}_s|  
R\sin\theta \cos\left(\phi_{{\bf k}_\parallel+{\bf k}_s}-\phi 
\right),
\end{equation} 
where $\phi_{\bf u}$ is the polar
angle of the vector ${\bf u}$ in the plane containing $\Lambda^*$ ($\Lambda$).
The only difference with respect to the case of a 2D periodicity in 3D,
which has been treated by Kambe \cite{Kam3,Kam2,Kam1} is
merely in that the values of 
$\phi_{{\bf k}_\parallel+{\bf k}_s}$ are no longer
from the interval $[0,2\pi)$ but are restricted
to either $0$ or $\pi$.
According to Eq. (\ref{il1int}), one has  
\begin{equation}  
I_L^{(1)}= i^{m-|m|} N_{l|m|} \sum_{n=0}^\infty  
\frac{1}{2^{2n} n!}  
\Gamma\left(-n, e^{-\pi i} 
   \frac{K_{\perp s}^2 \eta }{2}\right) K_{\perp s}^{2n}
   \left( \lim_{R\rightarrow 0}   
\frac{R^{2n} I_\Omega^n}{j_{l}(\sigma R)} \right),
\end{equation} 
where $I_\Omega^n$ involves the
following angular integral ($d\Omega=\sin\theta d\theta d\phi$),
\begin{equation}
I_\Omega^n \equiv
\int_0^\pi \sin\theta d\theta\, P^{|m|}_l (\cos\theta) (\cos\theta)^{2n}
\int_0^{2\pi}   e^{-im \phi +i
|{\bf k}_\parallel+{\bf k}_s| 
  R \sin\theta\cos\left(\phi_{{\bf k}_\parallel+{\bf k}_s}-\phi\right)}
  \, d\phi.
\label{iomn}
\end{equation}
Integrating first by $\phi$ one finds,
\begin{equation}
I_\Omega^n = 2\pi i^{|m|} e^{-im \phi_{{\bf k}_\parallel+{\bf k}_s}}
 \int_0^\pi  P^{|m|}_l (\cos\theta) (\cos\theta)^{2n}
  J_{|m|}( |{\bf k}_\parallel+{\bf k}_s| R \sin\theta )\, 
  \sin\theta d\theta,
\label{inom}
\end{equation}
where we have used that 
\begin{equation}
\int_0^{2\pi}   e^{-im \phi +i z
\cos(\phi_0 -\phi)}\, d\phi =2\pi i^{|m|} e^{-im \phi_0}
J_{|m|}( z ).
\end{equation}
The latter identity can be derived from 
(cf. Eqs. (9.1.44-45) of Ref. \cite{AS})
\begin{equation}
e^{iz\cos\phi} =\sum_{l=-\infty}^\infty i^{|l|} J_{|l|}(z) e^{il \phi}.
\label{eiex}
\end{equation}
Since $J_{\nu}(z)$ is an entire function of $z$, 
the Bessel function in Eq. (\ref{inom}) can be expanded  into power 
series of its argument (see Eq. (9.1.10) of Ref. \cite{AS}),
\begin{equation}
J_{|m|}(z)=\left(\frac{z}{2}\right)^{|m|} \sum_{j=0}^\infty
\frac{ \left(-z^2/4\right)^j }{j! (|m|+j)!}\cdot
\label{Jlex}
\end{equation}
Afterward Eq. (\ref{inom}) becomes
one obtains
\begin{equation}
I_\Omega^n = 2 \pi i^{|m|} e^{-im \phi_{{\bf k}_\parallel+{\bf k}_s}}
\sum_{j=0}^\infty (-1)^j 
\frac{[|{\bf k}_\parallel+{\bf k}_s| R]^{|m|+2j} }{2^{|m|+2j} j!
(|m|+j)!} \, I_\theta^j,
\end{equation}
where
\begin{equation}
I_\theta^j 
\equiv \int_0^\pi  (\cos\theta)^{2n}
  ( \sin\theta )^{|m|+2j+1} \, P^{|m|}_l (\cos\theta)  d\theta
= \int_{-1}^1 x^{2n} ( 1-x^2 )^{j+|m|/2}  P^{|m|}_l(x) dx.
\label{intheta}
\end{equation} 

It will turn out that, in addition to the general property (\ref{3ddgenpr}), 
each term $D_{lm}^{(j)}\equiv 0$, $j=1,2,3$, for $l-|m|$ {\em odd}.
In particular, one can show that the integral (\ref{intheta}) vanishes 
unless $l-|m|$ is {\em even}, i.e.,
\begin{equation}
D_{lm}^{(1)}\equiv 0, \hspace*{1.8cm} \mbox{$l-|m|$ odd}.
\label{3dd1genpr}
\end{equation}
Indeed,
\begin{equation}
P^{|m|}_l(\cos\theta)=(-1)^{l-|m|} P^{|m|}_l[\cos(\pi-\theta)].
\end{equation}
Now, if  $F$ is a real function such that
$F(\theta)=F(\pi-\theta)$, then for $l-|m|$ even one finds
\begin{equation}
\int_0^\pi F(\theta)P^{|m|}_l(\cos\theta) \, d\theta=
2 \int_0^{\pi/2}
F(\theta) P^{|m|}_l(\cos\theta)\, d\theta.
\end{equation}
On the other hand  for $l-|m|$  odd the integral vanishes.
Since in our case 
$$F(\theta)\equiv (\cos\theta)^{2n}
  ( \sin\theta )^{|m|+2j+1}=F(\pi-\theta),$$ 
it follows that $I_\theta^j \equiv 0$ for $l-|m|$ odd.

For $l-|m|$ even
\begin{equation}
\lim_{R\rightarrow 0}  
\frac{R^{2n} I_\Omega^n}{j_{l}(\sigma R)}\rightarrow 0,
\hspace*{1.5cm} j+ n > \frac{l-|m|}{2}\cdot
\end{equation}
Therefore, one only needs to investigate the case 
\begin{equation}
j+ n \leq  \frac{l-|m|}{2}\cdot
\label{jncon}
\end{equation}
Upon expanding $(1-x^2)^{j}$ in (\ref{intheta}) according to binomial 
theorem, $I_\theta^j$ is rewritten  as
\begin{equation}
I_\theta^j 
= \sum_{s=0}^j (-1)^s \left(
\begin{array}{c}
j\\
s
\end{array}\right)
\int_{-1}^1 x^{2(n+s)} ( 1-x^2 )^{|m|/2}  P^{|m|}_l(x) dx.
\label{ithj}
\end{equation}
According to Eq. (2.17.2.7) of Ref. \cite{PBM}, the integral 
\begin{equation}
\int_{-1}^1 x^{t} ( 1-x^2 )^{|m|/2}  P^{|m|}_l(x) dx 
\label{ithint} 
\end{equation} 
vanishes unless
\begin{equation}
|m|\leq l\leq t+|m|,
\end{equation}
or, in our case, unless
\begin{equation}
|m|\leq l\leq 2n+2s+|m| \Longleftrightarrow n+s \geq \frac{l-|m|}{2}\cdot
\label{nscon}
\end{equation}
Upon combining the conditions (\ref{jncon}) and (\ref{nscon}), it follows
that the only nonzero contribution to 
$I_\theta^j$ in the $R\rightarrow 0$ limit arises when simultaneously
\begin{equation}
s=j\hspace*{0.5cm}\mbox{and}\hspace*{0.5cm}  n +j= \frac{l-|m|}{2}\cdot
\label{jid}
\end{equation}
In the latter case, Eq. (2.17.2.6) of Ref. \cite{PBM} implies
\begin{eqnarray}
\lefteqn{ I_\theta^j =  (-1)^j
\frac{
2^{l+1}(2n+2j)!(l+|m|)![n+j+(l+|m|)/2]!}
{[n+j-(l-|m|)/2]!(2n+2j+l+|m|+1)!(l-|m|)!}}\hspace*{6cm}\nonumber\\
&& =  (-1)^{j}
\frac{
2^{l+1}l!(l+|m|)!}
{(2l+1)!}\cdot
\end{eqnarray}
Here in the last equation we have substituted for $n+j$ according
to Eq. (\ref{jid}). (Note in passing that 
Eq. (2.17.2.6) of Ref. \cite{PBM} differs by a factor
$(-1)^m/2$ compared to  Eq. (7.132.5) of Ref. \cite{GR} due to a 
slightly different definition of associated Legendre functions.) 
The constraints (\ref{jid}) imply  
\begin{equation}
|m|+2j=l-2n,
\end{equation}
and
\begin{equation}
n\leq \frac{l-|m|}{2}\cdot
\end{equation}
Therefore, the sum over $n$ in $I_L^{(1)}$ becomes a finite sum.

Consequently, as $R\rightarrow 0$,
\begin{eqnarray}
 R^{2n} I_\Omega^n &\sim& 
  e^{-im \phi_{{\bf k}_\parallel+{\bf k}_s}}
 (-1)^{j} \frac{2\pi i^{|m|} [|{\bf k}_\parallel+{\bf k}_s|]^{l-2n} }
{2^{l-2n} [(l-|m|)/2-n]![(l+|m|)/2-n]!}   
\nonumber\\
&&
(-1)^{j}
\frac{
2^{l+1}l!(l+|m|)!}
{(2l+1)!}\,R^l  \nonumber\\
&=& e^{-im \phi_{{\bf k}_\parallel+{\bf k}_s}}
  \frac{2 \pi i^{|m|} [|{\bf k}_\parallel+{\bf k}_s|]^{l-2n} }
{2^{l-2n} [(l-|m|)/2-n]![(l+|m|)/2-n]!} 
\frac{
2^{l+1}l!(l+|m|)!}
{(2l+1)!}\,R^l.
\end{eqnarray}
Now (Eq. (10.1.2) of Ref. \cite{AS}) 
\begin{equation}
j_l(\sigma R)\rightarrow \frac{2^{l}l!}{(2l+1)!}(\sigma R)^l
\hspace*{2cm}(R\rightarrow 0).
\end{equation}
Therefore, in the limit $R\rightarrow 0$:
\begin{eqnarray}  
I_L^{(1)} &=& 2\sqrt{\pi}\, \frac{i^{m}}{2^l} 
\left[ (2l+1)(l+|m|)!(l-|m|)!\right]^{1/2}  
e^{-im\phi_{{\bf k}_\parallel+{\bf k}_s}}\nonumber\\ && \times 
\sum_{n=0}^{(l-|m|)/2} \Gamma\left(-n, e^{-\pi i} 
   \frac{K_{\perp s}^2 \eta }{2}\right)  
  \frac{[|{\bf k}_\parallel+{\bf k}_s|/\sigma]^{l-2n} 
  [K_{\perp s}/\sigma]^{2n} } 
{n![(l-|m|)/2-n]![(l+|m|)/2-n]!}\cdot 
\end{eqnarray} 
Consequently, for $l-|m|$ {\em even}, 
\begin{eqnarray}  
D_{lm}^{(1)}(\sigma,{\bf k}_\parallel)  
&=&  - \frac{1}{4\pi v_0} I_L^{(1)}
\nonumber\\ 
&=& - \frac{1}{2 \sqrt{\pi} v_0} \frac{i^{m}}{2^l} 
\left[ (2l+1)(l+|m|)!(l-|m|)!\right]^{1/2} 
\sum_{{\bf k}_s\in\Lambda^*} e^{-im 
\phi_{{\bf k}_\parallel+{\bf k}_s}} \nonumber\\ 
&& \times 
\sum_{n=0}^{(l-|m|)/2} \Gamma\left(-n, e^{-\pi i} 
   \frac{K_{\perp s}^2 \eta }{2}\right)  
  \frac{[|{\bf k}_\parallel+{\bf k}_s|/\sigma]^{l-2n} 
  [K_{\perp s}/\sigma]^{2n} } 
{n![(l-|m|)/2-n]![(l+|m|)/2-n]!}, 
\nonumber\\
\label{dl1f1in3} 
\end{eqnarray} 
whereas [Eq. (\ref{3dd1genpr})]
\begin{equation}
D_{lm}^{(1)}\equiv 0, \hspace*{1.8cm} \mbox{$l-|m|$ odd}.
\end{equation}

\subsubsection{Complementary cases}
For $d_\Lambda=2$ and $d=3$, provided that
lattice plane is perpendicular to the $z$-axis, one can repeat most of the
steps presented here, the only change being $c=1/2$ instead of $c=1$,
and thereby arriving at the Kambe's expression \cite{Kam2,Pen,Moc}
\begin{eqnarray}  
D_{lm}^{(1)}(\sigma,{\bf k}_\parallel)  &=& - \frac{1}{2v_0 \sqrt{2\pi} } I_L^{(1)}
\nonumber\\ 
&=& - \frac{1}{\sigma v_0} \frac{i^{-m+1}}{2^l} 
\left[ (2l+1)(l+|m|)!(l-|m|)!\right]^{1/2} \sum_{{\bf k}_s\in\Lambda^*} e^{-im 
\phi_{{\bf k}_\parallel+{\bf k}_s}}
\nonumber\\ 
&& \times 
\sum_{n=0}^{(l-|m|)/2} \Gamma\left(1/2-n, e^{-\pi i} 
   \frac{K_{\perp s}^2 \eta }{2}\right)  
  \frac{[|{\bf k}_\parallel+{\bf k}_s|/\sigma]^{l-2n} [K_{\perp s}/\sigma]^{2n-1} } 
{n![(l-|m|)/2-n]![(l+|m|)/2-n]!}\cdot
 \nonumber\\ 
\label{dl1f2in3} 
\end{eqnarray}

For $d_\Lambda=1$ and $d=2$ one then finds \cite{Mol}
\begin{eqnarray}
\lefteqn{ D_{l}^{(1)} (\sigma,{\bf k}_\parallel)  
= - \frac{i^{|l|+1}|l|!}{\sqrt{2} \sigma v_0}\sum_{{\bf k}_s\in\Lambda^*}  
\sum_{n=0}^{[|l|/2]}\frac{1}{2^{2n} n!} \,\Gamma\left(1/2-n, e^{-\pi i}   
\frac{K_{\perp s}^2 \eta }{2}\right)} 
\nonumber\\
&& 
\times\frac{[|{\bf k}_\parallel+{\bf k}_s|/\sigma]^{|l|-2n} 
[K_{\perp s}/\sigma]^{2n-1}}{(|l|-2n)!}
\left\{
\begin{array}{cc}
\exp\left[-i(|l|-2n)\phi_{{\bf k}_\parallel+{\bf k}_s}\right], & l\geq 0\\
\exp\left[i(|l|-2n)\phi_{{\bf k}_\parallel+{\bf k}_s}\right], &   l< 0,
\end{array}\right.
\label{dl1f1in2}
\end{eqnarray}
where $v_0$ is now the length of the primitive cell of 
$\Lambda$ and $[|l|/2]$ stands for the integral part 
of $|l|/2$.

\subsubsection{Convergence and recurrence relations}
In virtue of the asymptotic behavior
\begin{equation}
\Gamma(a,z)\sim z^{a-1} e^{-z}
\hspace*{1.5cm} \mbox{as }
z\rightarrow\infty,\, |\mbox{arg }z|<3\pi/2  
\label{gmaz}
\end{equation}
(Eq. (6.5.32) of Ref. \cite{AS}) it is straightforward to verify 
that convergence of the series 
on the r.h.s. of expressions (\ref{dl1f1in3}),
(\ref{dl1f2in3}), (\ref{dl1f1in2}) is exponential
for sufficiently large $K_{\perp s}$.
One can verify that the lattice sums $D_{L}^{(1)}$ are {\em dimensionless} for $d=2$
whereas for $d=3$ the lattice sums $D_{L}^{(1)}$  have dimension $[1/\mbox{length}]$
[see discussion below Eq. (\ref{dstrco1})].
From the computational point of view, 
the incomplete gamma function in final expressions (\ref{dl1f1in3}),
(\ref{dl1f2in3}), (\ref{dl1f1in2}) can be derived  
successively by the recurrence formula \cite{Kam2,Pen}
\begin{equation} 
b\Gamma(b,x)=\Gamma(b+1,x) -x^b e^{-x} 
\label{igmfrec} 
\end{equation} 
from the value for $n=0$:
\begin{equation} 
\Gamma(1/2,x)=\left\{ 
\begin{array}{cc} 
\sqrt{\pi}-2\int_0^{\sqrt{x}} e^{-t^2}\, 
dt=\sqrt{\pi}\,\mbox{erfc}\,  
(\sqrt{x}), & \arg x=0 \\ 
\sqrt{\pi}\pm 2i \int_0^{\sqrt{-x}} e^{t^2}\, dt, & \arg x=\mp\pi. 
\end{array}\right. 
\label{igmferf} 
\end{equation}

\subsection{Calculation of $D_{L}^{(2)}$}
\label{sec:lyardl2} 

\subsubsection{General part} 
As it has been alluded to above, the contribution $D_{L}^{(2)}$ 
derives from the sum over direct lattice $\Lambda$ in 
the Ewald representation of $G_{0\Lambda}$, but with the term 
 ${\bf r}_n= 0$ {\em excluded}.
According to Eq. (\ref{dlcalc}) taken in combination
with the representations  (\ref{31grfr}), (\ref{32grfr}), (\ref{21grfr})  
of the quasi-periodic Green's functions,
\begin{equation}
D_{L}^{(2)} = - \frac{(2\pi)^{c}}{8\pi^2} I_L^{(2)},
\label{d2fcod1}
\end{equation}
where
\begin{equation} 
I_L^{(2)}=\int_{1/\eta}^{\infty} 
\zeta^{-c}\left( \lim_{R\rightarrow 0} \frac{1}{{\cal J}_{|l|}(\sigma R)} 
\oint  {\cal Y}_L^*(\hat{{\bf R}})  e^{\frac{1}{2}\, \left[ \sigma^2 /\zeta - 
 ({\bf R}+{\bf r}_s)^2\zeta\right] }\, 
\, d\Omega_{\bf R}\right) \, d\zeta.
\nonumber
\end{equation} 
Here $c=1/2$ for $d=3$ and $c=1$ for $d=2$.
Using the  plane-wave expansion (\ref{gplw}), which is also valid for 
complex arguments (see, e.g., Appendix 1 of Ref. \cite{Kam3} or
Appendix A of review by Tong \cite{Tong}), 
\begin{equation} 
e^{ -({\bf R}+{\bf r}_s)^2\zeta/2}=e^{-(R^2+r_s^2)\zeta/2} e^{i(i\zeta{\bf r}_s)\cdot{\bf R}}
= A e^{-(R^2+r_s^2)\zeta/2} \sum_{L} i^{|l|} {\cal J}_{|l|}(i\zeta r_s R)\, 
{\cal Y}_L^*(\hat{{\bf r}}_s) {\cal Y}_L(\hat{{\bf R}}). 
\label{exind2} 
\end{equation} 
The absolute value $|l|$ here is only relevant for $d=2$ and provided that 
the range of angular momenta is taken to be $-\infty <l<\infty$.

Substituting (\ref{exind2}) back into equation for $I_L^{(2)}$,
taking the limit (\ref{dlcalc}), and using that  in any dimension
\begin{equation} 
\frac{{\cal J}_l(az)}{{\cal J}_l(bz)}\rightarrow \frac{a^l}{b^l} 
\hspace*{1.5cm} \mbox{as } z\rightarrow 0,
\label{jlabp} 
\end{equation}   
\begin{equation} 
I_L^{(2)}= A i^{|l|} {\cal Y}_L^*(\hat{{\bf r}}_s) 
    \frac{(i r_s)^{|l|}}{\sigma ^{|l|}}\, 
\int_{1/\eta}^{\infty} \zeta^{|l|-c} \,   e^{\frac{1}{2}\, 
\left(\sigma^2 /\zeta -                            
 {\bf r}_s^2\zeta\right) }\, d\zeta. 
\nonumber
\end{equation} 
The substitution  
\begin{equation} 
\zeta=\frac{\sigma^2}{2u},\hspace*{1cm} r_s^2\zeta=\frac{\sigma^2r_s^2}{2u},
\hspace*{1cm} d\zeta =-\frac{\sigma^2}{2u^2}\,du,\hspace*{1cm} \alpha=\frac{\sigma^2\eta}{2}\cdot 
\label{subind2} 
\end{equation} 
then yields 
\begin{equation} 
I_L^{(2)}= A (-1)^{|l|} 2^{-|l|-1+c} \sigma^{2-2c} 
(\sigma r_s)^{|l|}  {\cal Y}_L^*(\hat{{\bf r}}_s) \int_{0}^{\alpha} u^{c-|l|-2} 
e^{ u - \sigma^2r_s^2/(4u) }\, du . 
\label{i2lmr} 
\end{equation} 
To this end, the formula for  $D_{L}^{(2)}$ is valid 
in any quasi-periodic case.

\subsubsection{Quasi-periodic cases in 3D} 
In 3D one has $c=1/2$ for both 1D and 2D periodicities.
Eq. (\ref{i2lmr}) then yields
\begin{equation}
I_L^{(2)}= 
4\pi (-1)^l 2^{-l-1/2} \sigma (\sigma r_s)^l Y_L^*(\hat{{\bf r}}_s) 
\int_{0}^{\alpha} u^{-l-3/2} e^{ u - \sigma^2r_s^2/(4u) }\, du .
\label{i2lmr3d}
\end{equation}
Now, using representations  (\ref{31grfr}), (\ref{32grfr}),
\begin{equation}
D_{L}^{(2)}=-\frac{(-1)^l\sigma}{2^l\sqrt{4\pi}} 
    \sum_{{\bf r}_s\in\Lambda}{}^{'} 
e^{-i{\bf k}_\parallel \cdot{\bf r}_s}
(\sigma r_s)^l Y_L^*(\hat{{\bf r}}_s) 
\int_{0}^{\alpha} u^{-l-3/2} e^{ u - \sigma^2r_s^2/(4u) }\, du,
\end{equation}
where prime in $\sum{}^{'}$ indicates as usual
that  the term with $r_s=0$ is omitted. Since, for $l+m$ {\em odd},
\begin{equation}
Y_{lm}(\pi/2,\phi)\equiv 0,
\label{diodpr}
\end{equation} 
\begin{equation}
D_{L}^{(2)}\equiv 0,\hspace*{2cm} l+m \,\,\mbox{ odd}.
\label{3dd2genpr}
\end{equation}
(We recall here that the periodicity 
direction(s) has been assumed to be perpendicular to the $z$-axis.)
This confirms [see Eq. (\ref{3dd1genpr})]
 that,
 in addition to the general property (\ref{3ddgenpr}), 
each term $D_{lm}^{(j)}\equiv 0$, $j=1,2,3$, for $l-|m|$ {\em odd}.
In the remaining cases for $l+m$ {\em even} \cite{Kam3,Kam2,Pen}
\begin{equation}
Y_{lm}^* (\pi/2,\phi)= (-1)^{(m-|m|)/2} \frac{(-1)^{(l+|m|)/2}}{2^{l} 
\sqrt{4\pi}}\,
\frac{[(2l+1)(l-|m|)!(l+|m|)!]^{1/2}}{[(l-|m|)/2]![(l+|m|)/2]!}\, 
e^{-im\phi}.
\label{diodpr2}
\end{equation}
Therefore,
\begin{eqnarray}
\lefteqn{D_{lm}^{(2)} (\sigma,{\bf k}_\parallel) =-\frac{\sigma}{4\pi}
\frac{(-1)^l(-1)^{(l+m)/2}}{2^{2l}}\, 
 \frac{[(2l+1)(l-|m|)!(l+|m|)!]^{1/2}}{[(l-|m|)/2]![(l+|m|)/2]!}
}\nonumber\\
&& \times
    \sum_{{\bf r}_s\in\Lambda}{}^{'} 
e^{-i{\bf k}_\parallel \cdot{\bf r}_s-im\phi_{{\bf r}_s}}
(\sigma r_s)^l 
\int_{0}^{\alpha} u^{-l-3/2} \exp\left[ u - \frac{\sigma^2r_s^2}{4u}\right]\, du,
\label{dl2f12in3}
\end{eqnarray}
where $\alpha=\sigma^2\eta/2$ [see Eq. (\ref{subind2})].
Hence, the respective expressions for $D_{lm}^{(2)}$ for $d_\Lambda=1$, $d=3$
and for $d_\Lambda=2$, $d=3$ \cite{Kam2,Pen,Moc} are formally identical
(provided that periodicity 
direction(s) is (are) perpendicular to the $z$-axis), 
the only difference  being in the lattice dimension.

\subsubsection{Complementary case of a 1D periodicity in 2D} 
In the remaining case for $d_\Lambda=1$ and $d=2$, in which case
$c=1$, one finds by a slight modification of the preceding derivation \cite{Mol}
\begin{eqnarray}
D_{l}^{(2)}(\sigma,{\bf k}_\parallel) 
&=&-\frac{(-1)^{|l|}}{2^{|l|+1}\sqrt{2\pi}} \,   
\sum_{{\bf r}_s\in\Lambda}{}^{'} e^{-i{\bf k}_\parallel 
\cdot{\bf r}_s}(\sigma r_s)^{|l|} e^{-il\phi_{{\bf r}_s}}
\nonumber\\
&& \times \int_{0}^{\alpha} u^{-|l|-1} 
\exp\left( u - \frac{\sigma^2r_s^2}{4u}\right)\, du.
\label{dl2f1in2}
\end{eqnarray}

\subsubsection{Convergence and recurrence relations}
Similarly as in the case of $D_{L}^{(1)}$,  convergence of the series 
for $D_{L}^{(2)}$ on the r.h.s. of expressions (\ref{dl2f12in3}),
 (\ref{dl2f1in2}) is exponential
for sufficiently large $r_s$. Indeed,  integrals
\begin{equation}
U_{|l|} = \int_{0}^{\alpha} u^{c-|l|-2} e^{ u - \sigma^2 r_s^2/(4u) }\, du
\label{ulint}
\end{equation}
are finite integrals. Now, for $|\sigma r_s|> |l|+2-c$, the integrand 
is monotonically increasing from zero to 
$(\sigma^2\eta/2)^{c-|l|-2} e^{\sigma^2\eta/2 - r_s^2/(2\eta)}$ on the 
integration interval. Therefore
\begin{equation}
|(\sigma r_s)^{|l|}  U_{|l|}| <  \sigma^{2(c-1)-|l|} (\eta/2)^{c-|l|-1}
 r_s^{|l|}  e^{\sigma^2\eta/2 - r_s^2/(2\eta)}.
\nonumber
\end{equation}

From the computational point of view, 
the integral on the r.h.s. of the resulting expressions
can easily be performed by 
a simple recurrence. Indeed, 
 knowing the values of $U_{0}$ and $U_{1}$, the respective integrals $U_{|l|}$
 can be determined using  recursion relation
\begin{equation}
\left(\frac{\sigma r_s}{2}\right)^2 U_{|l|+1}
= (|l|+1-c)U_{|l|} -U_{|l|-1} +
\alpha^{-|l|-1+c} e^{\alpha  - \sigma^2r_s^2/(4\alpha)}.
\label{dl2inrecr}
\end{equation}
The recurrence here follows, as suggested by Kambe (see Appendix 2 of Ref. \cite{Kam3}),
from a simple integration by parts.

As a consistency check, note that for $d=2$ the lattice 
sums $D_{L}^{(2)}$ are {\em dimensionless}, 
whereas for $d=3$ the lattice sums $D_{L}^{(2)}$  
have dimension $[1/\mbox{length}]$
[see discussion below Eq. (\ref{dstrco1})].

\subsection{Calculation of $D_{L}^{(3)}$} 
\label{sec:lyardl3} 

\subsubsection{General part} 
According to Eq. (\ref{dl3ex}), the only nonzero term is $D_{00}^{(3)}$,
or, for the sake of notation, $D_{0}^{(3)}$. 
In 3D, for both 1D and 2D periodicities, 
the $D_{0}^{(3)}$ term is calculated as the limit  
\begin{equation} 
D_{0}^{(3)}
 = \lim_{R\rightarrow 0} \frac{1}{{\cal J}_0(\sigma R){\cal Y}_0} 
\left[ -\frac{1}{4\pi^2} \left(\frac{\pi}{2} \right)^{1/2}  
 \int_{1/\eta}^{\infty} \zeta^{-1/2} e^{\frac{1}{2}\, 
\left(\sigma^2 /\zeta - {\bf R}^2\zeta\right) }\, 
d\zeta - G_0^p(\sigma,{\bf R})\right].
\label{dl3in3clc} 
\end{equation} 
Similarly, in 2D case
\begin{equation} 
D_{0}^{(3)} 
= \lim_{R\rightarrow 0} \frac{1}{{\cal J}_0(\sigma R){\cal Y}_0} 
\left[- \frac{1}{4\pi}   \int_{1/\eta}^{\infty} \zeta^{-1} e^{\frac{1}{2}\, 
\left( \sigma^2 /\zeta -                             
{\bf R}^2\zeta\right)}\, d\zeta - G_0^p(\sigma,{\bf R})\right]. 
\label{dl3in2clc} 
\end{equation} 
It is reminded here that $G_0^p(\sigma,{\bf R})$ in Eqs. (\ref{dl3in3clc}) and (\ref{dl3in2clc}) 
is the corresponding singular, or principal-value, part of $G_0^+(\sigma,{\bf R})$ 
[see Eq. (\ref{singp}) above].

In any dimension ${\cal Y}_0 =A^{-1/2}$ [Eq. (\ref{yps0})], where 
$A$ is given by Eq. (\ref{Adef}), and 
(see Eqs. (9.1.12), (10.1.11) of Ref. \cite{AS}) 
\begin{equation} 
{\cal J}_0(\sigma R)\rightarrow 1 \hspace*{1cm} (R\rightarrow 0). 
\label{j0lim}
\end{equation} 
Hence, calculation of $D_{0}^{(3)}$ requires to
 perform integral 
\begin{equation} 
I^{(3)}=\int_{1/\eta}^{\infty} \zeta^{-c} e^{\frac{1}{2}\, 
\left(\sigma^2 /\zeta - {\bf R}^2\zeta\right)}\, d\zeta 
\nonumber
\end{equation} 
for either $c=1/2$ ($d=3$) or $c=1$ ($d=2$). 
Expanding $e^{\sigma^2 /(2\zeta)}$ into power series and integrating
 term by term yields 
\begin{equation} 
I^{(3)}= \sum_{n=0}^\infty  
\frac{\sigma^{2n}}{2^n n!}\int_{1/\eta}^{\infty} 
\zeta^{-c-n} e^{- {\bf R}^2\zeta/2 }\, d\zeta. 
\nonumber
\end{equation} 
Since all terms in the sum are positive, 
the exchange of integration and summation is 
justified whenever one shows that one of the sides 
exists (this will be shown later on). 
Upon substituting $\zeta=2t/R^2$, 
\begin{eqnarray}
I^{(3)} &=& (R^2/2)^{c-1} \sum_{n=0}^\infty   
\frac{(\sigma^2 R^2)^{n}}{2^{2n} n!}\int_{R^2/(2\eta)}^{\infty} 
t^{-c-n} e^{- t }\, d t\nonumber\\
&=& (R^2/2)^{c-1} \sum_{n=0}^\infty   
\frac{(\sigma R)^{2n}}{2^{2n} n!} \, \Gamma[-n-c+1,R^2/(2\eta)], 
\nonumber
\end{eqnarray}
where as usual $\Gamma(a,x)$ is an incomplete gamma function 
(see Eq. (6.5.3) of Ref. \cite{AS}). 
According to Eqs. (\ref{dl3in3clc}) and 
(\ref{dl3in2clc}),   we are only 
interested in the limit $R\rightarrow 0$. 
For $n> 1-c$, which given  $c=1/2,$ $1$ translates into $n\geq 1$, 
one can integrate by parts, yielding 
\begin{equation} 
\int_{R^2/(2\eta)}^{\infty} t^{-c-n} e^{- t }\, d t 
= \frac{1}{n+c-1}  \left( \frac{R^2}{2\eta}\right)^{-c-n+1} e^{-R^2/(2\eta)} 
+ {\cal O}\left[R^{-2(c+n-2)}\right]. 
\nonumber
\end{equation} 
Therefore, 
\begin{equation} 
I^{(3)}= \sum_{n=1}^\infty \frac{1}{n+c-1} 
 \frac{(\sigma^2 \eta)^{n}}{2^{n} n!} +\lim_{R\rightarrow 0} 
(R^2/2)^{c-1} \int_{R^2/(2\eta)}^{\infty} t^{-c} e^{- t }\, d t. 
\label{i3lint} 
\end{equation} 
A further treatment differs in different dimensions.

\subsubsection{Quasi-periodic cases in 3D} 
In 3D one has $c=1/2$ for both 1D and 2D periodicities. Hence,
\begin{eqnarray}
\int_{R^2/(2\eta)}^{\infty} t^{-c} e^{- t }\, d t
 &=& \left( \int_{0}^\infty -  \int_0^{R^2/(2\eta)}\right) t^{-1/2} 
e^{- t }\, d t 
\nonumber\\
&=& 
\Gamma(1/2) - 2 \frac{R}{(2\eta)^{1/2}} e^{-R^2/(2\eta)} + {\cal O}(R^{3}),
\end{eqnarray}
where $\Gamma(1/2)=\sqrt{\pi}$.
This asymptotic is consistent with that obtained by 
writing
\begin{equation} 
\Gamma[1/2,R^2/(2\eta)]=\int_{R^2/(2\eta)}^{\infty} 
t^{-1/2} e^{- t }\, d t= \sqrt{\pi} \, \mbox{erfc}\, 
\left( R/\sqrt{2\eta}\right) 
\label{gamma0n}
\end{equation} 
(see Eq. (6.5.17) 
of Ref. \cite{AS}) and using that 
\begin{equation} 
\mbox{erfc}\, z = 1-\frac{2z}{\sqrt{\pi}} +{\cal O}(z^2) 
\hspace*{1.5cm} \mbox{as }z\rightarrow 0
\label{erfcas}
\end{equation} 
(see Eq. (7.2.4) of Ref. \cite{AS}). 
Therefore, 
\begin{equation} 
I^{(3)}= - \frac{\sqrt{2\pi} }{R}+\frac{\sigma}{\sqrt{2}} 
\sum_{n=0}^\infty  \frac{(\sigma^2\eta/2)^{n-1/2}}{n!(n-1/2)} 
+ {\cal O}(R^{2}) \hspace*{1.2cm} \mbox{as }R\rightarrow 0.
\nonumber
\end{equation} 
Collecting everything together back to 
Eq. (\ref{dl3in3clc}), the first term in $I^{(3)}$ 
cancels against $G_0^p(\sigma,{\bf R})$ leaving behind
\begin{equation} 
D^{(3)}_{lm}  (\sigma) = -\frac{\sigma}{4\pi}  
\sum_{n=0}^\infty \frac{(\sigma^2\eta/2)^{n-1/2}}{n!(n-1/2)} 
 \, \delta_{lm,00}. 
\label{dl3f1in3} 
\end{equation}
It is emphasized here that the respective contributions $D^{(3)}_{lm}$
for a 1D periodicity in 3D and a 2D periodicity in 3D (see Refs. \cite{Kam2,Pen,Moc})
are identical.

\subsubsection{Complementary case of 1D periodicity in 2D}
For $d_\Lambda=1$ and $d=2$, in which case $c=1$,  one finds \cite{Mol}
\begin{equation}
D^{(3)}_l (\sigma)
=-\frac{1}{2\sqrt{2\pi}}\left[\gamma + \ln (\sigma^2\eta/2)  
+ \sum_{n=1}^\infty \frac{(\sigma^2\eta/2)^n}{n!n} \, \right] 
\delta_{l0}=\frac{1}{2\sqrt{2\pi}} \, \mbox{Ei} (\sigma^2\eta/2) \,\delta_{l0},
\label{dl3f1in2}
\end{equation}
where Ei is an exponential integral and 
$\gamma\approx 0.57721\, 56649$ is the Euler constant 
(see Eqs. (5.1.10) and (6.1.3) of Ref. \cite{AS}, respectively). 
Note that, for 
$\Lambda$ oriented along the $x$-axis, 
$D_{l}^{(j)}=D_{-l}^{(j)}$, $j=1,\,2$, 
and hence, $D_{l}=D_{-l}$, in accord with the fact 
that $G_{0\Lambda}$ only depends on 
$y$ via $|y|$ \cite{NMcP,YYo}.

\subsubsection{Convergence}
Unlike preceding cases, convergence 
of series on the r.h.s. of Eqs. (\ref{dl3f1in3}) and 
(\ref{dl3f1in2}) is even faster than exponential. This can easily be  verified 
by using Stirling's formula (Eq. (6.1.37) of Ref. \cite{AS})
\begin{equation}
(n+1)! \sim \sqrt{2\pi n}\, n^n e^{-n} \hspace*{1.5cm} \mbox{as }n\rightarrow\infty.
\label{strlng}
\end{equation}
Again, as a consistency check, note that for $d=2$ 
the lattice sums $D_{L}^{(3)}$ 
are {\em dimensionless}, 
whereas for $d=3$ the lattice sums $D_{L}^{(3)}$  
have dimension $[1/\mbox{length}]$
[see discussion below Eq. (\ref{dstrco1})]. It is reminded here that $\eta$
in the above formulae has dimension of $[\mbox{length}]^2$.

\section{Laplace equation}
\label{sec:pois}
The quasi-periodic solutions of the Laplace equation  
are used to describe potential flows 
in fluid dynamics between 
parallel planes and in rectangular channels \cite{Lnt}.
Indeed, a Green's function representing a point source and 
satisfying the respective 
von Neumann and Dirichlet boundary conditions on 
a flow channel walls can be written
as a sum and difference of an appropriate $G_{0\Lambda}$ 
(corresponding to 1D periodicity in 3D for the flow between parallel planes
and to 2D periodicity in 3D for the flow in a rectangular channel) 
taken at two different spatial points \cite{Lnt}.
Another important class of problems associated 
with the quasi-periodic solutions of Laplace equation
arises in various problems
in electrostatics and elastostatics \cite{QLK,PaS,GAu1}.

The relevant representations of $G_{0\Lambda}$ for
the Laplace equation can in principle
be obtained by taking the limit $\sigma \rightarrow 0$ in the resulting
expressions for the Helmholtz equation. 
In 3D, $h_{0}^{(1)}(\sigma |{\bf r}-{\bf r}'+{\bf r}_s|)$
in the Schl\"{o}milch  series (\ref{fsgfls}) exhibits a regular limit 
\begin{equation}
 h_{0}^{(1)}(\sigma |{\bf r}-{\bf r}'+{\bf r}_s|) \rightarrow 
       -\frac{i}{|{\bf r}-{\bf r}'+{\bf r}_s|}
\hspace*{1.5cm} \mbox{as }\sigma \rightarrow 0
\nonumber
\end{equation}
[see Eq. (\ref{h0z})] and corresponding 
 $G_0^+ (\sigma,R)$ goes smoothly to the free-space Green's function
 of 3D Laplace equation. However, 
$H_{0}^{(1)}(\sigma |{\bf r}-{\bf r}'+{\bf r}_s|)$ displays
a logarithmic singularity in the same limit 
(see Eqs. (9.1.3), (9.1.13) of Ref. \cite{AS}).
Hence $G_0^+ (\sigma,R)$ does not reduce to the free-space Green's function
 $G_0^+ (R) =  (1/2\pi) \ln R $ of 2D Laplace equation.
Surprisingly enough, in the case of dual and 
Ewald representations of $G_{0\Lambda}$ the 
logarithmic singularities cooperate in such a way that 
the limit $\sigma \rightarrow 0$
turns out to be regular even for $d=2$ (see below).

In the following, the limit $\sigma \rightarrow 0$ will be discussed in the case 
of dual and Ewald representations of $G_{0\Lambda}$, and in the case of lattice sums
$D_{00}$ in 3D. 
In taking the limit, both $G_{0\Lambda}$ and $D_{00}$ will be 
considered formally as functions of two independent variables $\sigma$ and
${\bf k}_\parallel$. A reason for doing so is, for instance, solving
an implicit equation (\ref{zps}). 
The limit ${\bf k}_\parallel \rightarrow 0$ will be,
if possible, considered afterwords.

Note that  the case of 1D periodicity 
in the $x$-direction in 2D is rather academic in what follows, since 
in the latter case the Green's function can be calculated in a closed form \cite{Lnt}
\begin{equation} 
G_{0\Lambda}({\bf R})=\frac{1}{\pi} \ln\left[
2\left| \sin \frac{\pi}{a} (x+iy) \right| \right].
\end{equation}

\subsection{Dual representations}
\label{sec:lpld}
Since $K_{\perp n} \rightarrow i|{\bf k}_\parallel +{\bf k}_n|$ 
in the limit $\sigma \rightarrow 0$ [cf. Eq. (\ref{Kkn})],
in all quasi-periodic cases the
limit is established [see Eq. (\ref{lplim})] by replacing 
${\cal H}_{0}^{+} (K_{\perp n} |R_\perp|)$ in Eqs. 
 (\ref{31dual}), (\ref{32dual}) with
${\cal H}_{0}^{+} (i|{\bf k}_\parallel +{\bf k}_n| |R_\perp|)$.
However, for purely imaginary argument $ix$ with $x>0$ the Hankel functions
${\cal H}_{0}^{+} (ix)$ are related to modified Bessel functions 
of third kind (see Eqs. (9.6.4) and (10.2.15) of Ref. 
\cite{AS}). This results in rapidly decaying terms and exponential convergence.
As it has been alluded to earlier, absolute exponential convergence
of the dual representations can only be established under the assumption of
$R_\perp\neq 0$. 

However, the resulting dual representations
are singular in the limit
${\bf k}_\parallel \rightarrow 0$. 
Then $K_{\perp n} \rightarrow i k_n$ and the denominator
in Eqs. (\ref{31dual}), (\ref{32dual}) vanishes for ${\bf k}_n=0$.

\subsection{Ewald representations}
\label{sec:lplew}
The $\sigma \rightarrow 0$ limit can also be easily taken in the 
respective Ewald integral representations 
(\ref{31grfr}), (\ref{32grfr}), (\ref{21grfr}) of
free-space quasi-periodic Greens functions: simply substitute 
in the above expressions
$\phi_n\equiv 0$ and $K_{\perp n}^2=-|{\bf k}_\parallel+{\bf k}_n|^2$.
Using that the integrals in the series
over $\Lambda$ can be expressed in the $\sigma \rightarrow 0$ limit 
in terms of incomplete gamma functions
(Eq. (6.5.3) of Ref. \cite{AS}), one finds 
for the respective Ewald representations 
(\ref{31grfr}), (\ref{32grfr}), (\ref{21grfr}) 
 the following expressions:
\begin{eqnarray}
G_{0\Lambda}({\bf k}_\parallel,{\bf R}) &=& 
- \frac{1}{4\pi v_0} 
\sum_{{\bf k}_n\in\Lambda^*} 
e^{i({\bf k}_\parallel+{\bf k}_n)\cdot{\bf R}_\parallel}
  \int_{\eta}^{\infty}
\zeta^{-1} 
e^{- \frac{1}{2}\, (|{\bf k}_\parallel+{\bf k}_n|^2 \zeta + |R_\perp|^2/\zeta)}\,
d\zeta\nonumber\\
&&
-\frac{1}{4\pi^2} \left(\frac{\pi}{2} \right)^{1/2}
 \sum_{{\bf r}_s\in\Lambda} e^{-i{\bf k}_\parallel \cdot{\bf r}_s}
\frac{\sqrt{2}\, \Gamma[1/2,|{\bf R}+{\bf r}_s|^2/(2\eta)] }{|{\bf R}+{\bf r}_s|}
\label{31grfrl}
\end{eqnarray}
for 1D in 3D,
\begin{eqnarray}
G_{0\Lambda}({\bf k}_\parallel,{\bf R}) &=&
-\frac{1}{2\pi v_0} \left(\frac{\pi}{2} \right)^{1/2}
\sum_{{\bf k}_n\in\Lambda^*} 
   e^{i({\bf k}_\parallel+{\bf k}_n)\cdot{\bf R}_\parallel}
  \int_{\eta}^{\infty}  \zeta^{-1/2} 
e^{-\frac{1}{2}\, (|{\bf k}_\parallel+{\bf k}_n|^2 \zeta + |R_\perp|^2/\zeta)}\,
d\zeta\nonumber\\
&&
-\frac{1}{4\pi^2} \left(\frac{\pi}{2} \right)^{1/2}
     \sum_{{\bf r}_s\in\Lambda} 
      e^{-i{\bf k}_\parallel \cdot{\bf r}_s}
  \frac{\sqrt{2}\, \Gamma[1/2,|{\bf R}+{\bf r}_s|^2/(2\eta)] }{|{\bf R}+{\bf r}_s|}
\label{32grfrl}
\end{eqnarray}
for 2D in 3D and
\begin{eqnarray}
G_{0\Lambda}({\bf k}_\parallel,{\bf R}) 
&=& - \frac{1}{2\pi v_0} \left(\frac{\pi}{2} \right)^{1/2}
\sum_{{\bf k}_n\in\Lambda^*} 
   e^{i({\bf k}_\parallel+{\bf k}_n)\cdot{\bf R}_\parallel}
  \int_{\eta}^{\infty}
\zeta^{-1/2} 
 e^{- \frac{1}{2}\, (|{\bf k}_\parallel+{\bf k}_n|^2 \zeta+ |R_\perp|^2/\zeta)}\,
d\zeta\nonumber\\
&&
- \frac{1}{4\pi} \sum_{{\bf r}_s\in\Lambda} 
    e^{-i{\bf k}_\parallel \cdot{\bf r}_s}  \Gamma[0,|{\bf R}+{\bf r}_s|^2/(2\eta)]
\label{21grfrl}
\end{eqnarray}
for 1D in 2D. The Laplace limit of the respective Ewald integral representations 
then follows straightforwardly
by letting  ${\bf k}_\parallel \rightarrow 0$ in the above expressions
(\ref{31grfrl}), (\ref{32grfrl}), (\ref{21grfrl}).

Regarding convergence speed, in virtue of the asymptotic 
$\Gamma(a,z)\sim z^{a-1} e^{-z}$ for
$z\rightarrow\infty$, $|\mbox{arg }z|<3\pi/2$  (Eq. (6.5.32) of Ref. \cite{AS}) 
the respective Ewald representations remain to be exponentially
convergent in the $\sigma \rightarrow 0$, ${\bf k}_\parallel \rightarrow 0$  limit.
Note in passing that
$\Gamma[1/2,|{\bf R}+{\bf r}_s|^2/(2\eta)]$ in Eqs. (\ref{31grfrl}), 
(\ref{32grfrl}) 
can be expressed via error function [Eq. (\ref{gamma0n})] as
\begin{equation}
\Gamma[1/2,|{\bf R}+{\bf r}_s|^2/(2\eta)] = \sqrt{\pi}\, 
   \mbox{erfc} (|{\bf R}+{\bf r}_s|/\sqrt{2\eta})
\end{equation}
(Eq. (6.5.17) of Ref. \cite{AS}).
An alternative exponentially convergent series for $G_{0\Lambda}$ 
for 1D periodicity in 3D in the Laplace case has also been obtained earlier
by Linton (see series in Eq. (3.26) of Ref. \cite{Lnt}). 
However, our expressions have been derived
without any artificial
regularization in the form of a  convergence ensuring logarithmically 
divergent series (cf. Ref. \cite{Lnt}).

Additionally, absolute exponential convergence
of the respective Ewald representations 
can also be established for
$R_\perp = 0$, which for numerous alternative representations 
of $G_{0\Lambda}$ in fluid dynamics provides a problem \cite{Lnt}.
Again, the case $R_\perp = 0$
can only be attained when ${\bf R}_\parallel \not\in\Lambda$.
Otherwise the
respective Ewald representations become singular.
In 3D this singularity is explicit, since for some ${\bf r}_s\in\Lambda$
the denominator $|{\bf R}+{\bf r}_s|$ vanishes. For 1D in 2D
one has $\Gamma[0,|{\bf R}+{\bf r}_s|^2/(2\eta)] \rightarrow \Gamma(0)$ 
for some ${\bf r}_s\in\Lambda$,
and the singular behavior follows from the pole of 
the gamma function $\Gamma(z)$ for $z=0$, or more precisely 
from the asymptotic
(Eqs. (6.5.15), (5.1.11) of Ref. \cite{AS})
\begin{equation}
\Gamma (0,z) = E_1(z) \sim -\gamma - \ln z -\sum_{n=1}^\infty \frac{(-z)^n}{nn!}
\hspace*{1.5cm} \mbox{as } z\rightarrow 0,
\end{equation}
where $\gamma$ is the Euler constant 
(Eq. (6.1.3) of Ref. \cite{AS}).

As a final remark, note in passing that each of the above Ewald integral representations
(\ref{31grfrl}), (\ref{32grfrl}), (\ref{21grfrl}) can be regarded
as a one-parametric continuous spectrum of 
the representations for $G_{0\Lambda}$.
The corresponding dual representations of Sec. \ref{sec:lpld}
for ${\bf k}_\parallel \neq 0$  can be then recovered in the limit $\eta\rightarrow 0$.
Indeed, using Hobson's 
integral representation (\ref{hobs}), 
the integrals in the series over $\Lambda^*$
can be expressed in the limit $\eta\rightarrow 0$ in terms of the modified
Bessel functions of the third kind ${\cal K}_0$ ($K_0$ for $d-d_\Lambda=2$
and $K_{1/2}$ for $d-d_\Lambda=1$).

\subsection{Lattice sums in 3D for $l=0$}
\label{sec:lpldir}
In the case of lattice sums $D_{L}$, they 
are defined as expansion coefficients
of free-space quasi-periodic Greens functions in terms of 
regular cylindrical (in $2D$ or spherical (in $3D$) 
waves ${\cal J}_{l}(\sigma  R) {\cal Y}_L(\hat{{\bf R}})$
[see Eq. (\ref{dstrco}) above)]. Since unless $l=0$ 
one has ${\cal J}_{l}(\sigma  R)\rightarrow 0$
in the limit $\sigma \rightarrow 0$ [see Eq. (\ref{j0lim})], the lattice 
sums $D_{L}$ become singular in the limit for $l\neq 0$.
In the case of $l=0$ it is reminded here that $D_{00}$ 
($\gamma$ function of Karpeshina \cite{Kar,Krp})
determines the  energy operator 
in the one-particle theory of periodic point (zero range)
interactions 
 and that
the  $D_{00}$ determines the underlying 
spectrum according to Eq. (\ref{zps}).

From a mathematical point of view, upon substituting 
(\ref{h0z}) for ${\cal H}_0^+$
in (\ref{dstrcos}) and assuming for a while an {\em integer lattice},
the lattice sum $D_{L}$ for $l=0$ in 3D 
can be expressed in the limit $\sigma \rightarrow 0$
via {\em Epstein zeta function} \cite{Em1,Ep,GlZ}
\begin{equation}
 Z
\left|
\begin{array}{c}
0 \\
{\bf k}_\parallel
\end{array}
\right|(\chi,\nu)
\equiv
\sum_{{\bf r}_n\in\Lambda}{}^{'}  
\frac{e^{ i {\bf k}_\parallel\cdot {\bf r}_n} }
{|{\bf r}_n|^{\nu}}
\label{dstrcose}
\end{equation}
as 
\begin{equation}  
 D_{00}(0,{\bf k}_\parallel) =  - \frac{1}{\sqrt{4\pi}}
 \, Z
\left|
\begin{array}{c}
0 \\
{\bf k}_\parallel
\end{array}
\right|(\chi,1),
\label{dstrcosl} 
\end{equation}  
where we have used Eq. (\ref{yps0}) and [Eq. (\ref{CArel})] 
that $C\rightarrow 0$ as $\sigma\rightarrow 0$ in 3D.
In Eq. (\ref{dstrcose}) for $d_\Lambda=2$, $\chi$ is  
a positive quadratic form defined by the scalar
product of the basis vector of $\Lambda$,
 $\chi_{ij}={\bf e}_i\cdot{\bf e}_j$. (For $d_\Lambda=1$, all the
quadratic forms $\chi$ reduce to an absolute value).
The Epstein zeta function in Eq. (\ref{dstrcose})
converges absolutely for Re $\nu> d_\Lambda$ and 
can be analytically continued to an  entire
function in the complex variable $\nu$ unless ${\bf k}_\parallel \in\Lambda^*$,
in which case the Epstein zeta function possesses a
simple pole for $\nu=d_\Lambda$. 

Interestingly enough,
for $d_\Lambda=2$, $d=3$, $\nu=1$ the Epstein zeta function 
can be expressed in  a closed form
in terms of Jacobi theta functions \cite{Gla3}.
For a 1D periodicity in 3D with a period $a$ one then,
starting from Eq. (\ref{dstrcos0}), obtains
in the limit $\sigma \rightarrow 0$  
\begin{equation}  
 D_{00}(0,{\bf k}_\parallel) =  \frac{1}{\sqrt{4\pi} a}
\ln\left[2 - 2\cos({\bf k}_\parallel a) \right].
\label{dstrcos0l} 
\end{equation}  
The expression is obviously singular for ${\bf k}_\parallel \in\Lambda^*$,
in accordance with the singularity of the Epstein zeta function 
for $\nu=d_\Lambda=1$.

Regarding the Epstein zeta functions, note that 
one could have written dual representations
in the limit $\sigma \rightarrow 0$ for $R_\perp= 0$, $d=3$, and 
provided that ${\bf k}_\parallel\not\in\Lambda^*$, as
\begin{equation}
G_{0\Lambda}(\sigma,{\bf k}_\parallel,{\bf R}) =
- \frac{e^{i {\bf k}_\parallel\cdot {\bf R}_\parallel}}{2 v_0} 
 \, Z
\left|
\begin{array}{c}
{\bf k}_\parallel \\
{\bf R}_\parallel
\end{array}
\right|(\chi,1),
\label{32duall}
\end{equation}
where
\begin{equation}
Z
\left|
\begin{array}{c}
{\bf k}_\parallel \\
{\bf R}_\parallel
\end{array}
\right|(\chi,\nu) = \sum_{{\bf k}_n\in\Lambda^*}{}^{'}  
\frac{e^{i {\bf k}_n\cdot{\bf R}_\parallel}}
{|{\bf k}_\parallel +{\bf k}_n|^{\nu}}\cdot
\label{epst}
\end{equation}

\section{Discussion}
\label{sec:disc}

\subsection{The choice of the Ewald parameter $\eta$}
Each of the exponentially convergent Ewald representations (\ref{31grfr}), 
(\ref{32grfr}), (\ref{21grfr}) 
for $G_{0\Lambda}$ can be viewed as a one-parametric family of representations. 
A corresponding image-like series (\ref{lgrfdf}) and a
dual representation [Eqs. (\ref{31dual}), (\ref{32dual})]  
can be seen then as two ends of the one-parametric continuous spectrum of 
the representations for $G_{0\Lambda}$.
Obviously, by varying the point $\eta$, at which 
the integration is split, the convergence characteristics of the
representation can be altered.
In most cases, the value of the Ewald parameter $\eta$
is chosen to balance the convergence of respective 
$D_{L}^{(1)}$ and $D_{L}^{(2)}$ contributions.
This leads occasionally to the criticism that 
an arbitrary optimization parameter enters 
the evaluation of lattice sums. On contrary,
the invariance of $D_L$'s on the value of Ewald parameter
$\eta$ serves as a check of a correct numerical
implementation. The Ewald parameter $\eta$ can often be varied by several
orders of magnitude without affecting the results
in a wide frequency window. However, for some
range of $\eta$ values one can enter a numerically unstable
region: the respective 
$D_L^{(1)}$ and $D_L^{(2)}$ contributions have opposite
signs and similar magnitude, which is several orders
larger than the magnitude of resultant $D_L$. This instability can easily
be remedied by the choice of some other value of $\eta$, or
one can follow the recipe of Berry \cite{Be} and 
chose $\eta$ to depend on $\sigma$ and $l$, and thereby prevent numerical
instability completely.
Indeed, although the results presented here have been obtained 
by a uniform $l$-independent  choice of $\eta$, one can easily 
modify the above 
derivations to the case of $l$-dependent $\eta$ \cite{Be}.

\subsection{Numerical convergence}
\label{app:conv}  
Exponentially convergent representations of 
lattice sums summarized here
provide a significant advantage in terms 
of computational speed, while maintaining accuracy,
over alternative expressions of 
lattice sums. In the special case of a 1D periodicity in 2D 
this is demonstrated in Table I below.
In the case of 1D periodicity in 2D, even
with the latest progress due to Yasumoto and Yoshitomi \cite{YYo}, it took
$40$ seconds to compute $G_{0\Lambda}$ 
on SPARC workstation from lattice
sums with $14$ digits accuracy at a single point and frequency. This
was in striking contrast to the  calculation of exponentially convergent
lattice sums in the so-called bulk cases, i.e.,
when $G_{0\Lambda}$  is periodic in all space dimensions.
The lattice sums for an infinite 2D lattice in 2D \cite{OA} and
an infinite 3D  lattice in 3D \cite{HS}. The
respective convergence times (on a PC with Pentium II processor) for a set of
bulk 2D and 3D lattice sums  with $6$ digits accuracy are less than 
$\approx 0.03$ second (for a cut-off value of $l_{max}=20$) \cite{LM} 
and $\approx 0.8$ second (for a cut-off value of $l_{max}=6$) \cite{MS}.

\begin{center}
TABLE I. Quasi-periodic Green's function $G_{0\Lambda}$ 
for off-axis incidence at an angle
$\theta=\pi/8$ upon a 1D lattice
oriented along the $x$-axis in 2D with
$\lambda/v_0=0.23$. Here $v_0$ is the length of a period
(primitive lattice cell) along the $x$-axis. In rows
labeled by D, values of $G_{0\Lambda}$ obtained by a direct summation  of its dual
representation (spectral domain form) (taken from Tab. III of Ref. \cite{NMcP}).  
These data are compared against those in rows labeled by E, 
obtained by the (complete) Ewald-Kambe summation [Eqs. 
(\ref{dl1f1in2}), (\ref{dl2f1in2}), (\ref{dl3f1in2}) with the Ewald parameter $\eta=0.011$]. 
Data in the respective rows labeled by YY and NMcP are those  obtained by
 Yasumoto and Yoshitomi \cite{YYo} and Nicorovici and McPhedran \cite{NMcP} methods.  
\vspace*{0.5cm} \\
\begin{tabular} {|c|c|c|c|c|} \hline\hline
 & & & &  \\
   &   $x$  &  $y$      &  Re $G_{0\Lambda}$   &  Im $G_{0\Lambda}$ \\
\hline\hline
 & & & &  \\
D  &  $0.2$ &  $0.03$   &  $0.117120006144932$ &  $-0.108131857633201$ \\
E  &   &                &  $0.117120006144932$ &  $-0.108131857633206$\\ 
YY &   &                &  $0.117120006144941$ &  $-0.108131857633206$\\ 
NMcP  &   &             &  $0.117120006141860$ &  $-0.108131857633197$\\ \hline 
D  &  $0.2$ &  $0.003$  &  $0.115891895634567$ &  $-0.103497063599642$\\  
E  &   &                &  $0.115891895634565$ &  $-0.103497063599651$\\  
YY &   &                &  $0.115891895634577$ &  $-0.103497063599646$\\
NMcP  &   &             &  $0.115891895630095$ &  $-0.103497063599643$\\  \hline
D  &  $0.2$ &  $0.0003$ &  $0.115881138140449$ &  $-0.103450147416784$ \\ 
E       &   &           &  $0.115881138140448$ &  $-0.103450147416794$\\ 
YY &   &                &  $0.115881138140457$ &  $-0.103450147416788$\\ 
NMcP &   &              &  $0.115881138135960$ &  $-0.103450147416785$\\                                   
 \hline\hline
\end{tabular}
\end{center}
\vspace*{0.5cm} 
\noindent

The computational time to reproduce a value of $G_{0\Lambda}$ in 
Table I with accuracy of within $8\times 10^{-15}$ of that
obtained by a direct summation turns out to be $\approx 0.2$s, in
line with the respective $\approx 0.03$s and $\approx 0.8$s for convergence
time of a set of bulk 2D \cite{LM} and 3D lattice sums \cite{MS} with $6$ 
digits accuracy. This should be compared to $1232$s 
of Nicorovici and McPhedran \cite{NMcP},
or, to $40$s of Yasumoto and Yoshitomi \cite{YYo} (the computational times
have been taken from Ref. \cite{YYo}).
The exponentially convergent representation 
[Eqs. (\ref{dl1f1in2}), (\ref{dl2f1in2}), (\ref{dl3f1in2})] 
(i) can be implemented numerically more simply and (ii) converges roughly
$200$ times faster than the previous best representation \cite{YYo}. 
Of the cases tested, the
simplest case of a constant $\eta=0.011$ was chosen.

For a 2D periodicity in 3D, a comparison of the speed and accuracy
of exponentially convergent representation of 
lattice sums [Eqs. (\ref{dl1f2in3}), (\ref{dl2f12in3}), (\ref{dl3f1in3})]
with respect to  alternative expressions for $G_{0\Lambda}$ 
has been summarized in Ref. \cite{Moc}. Again, 
exponentially convergent representation of 
lattice sums turns out to be convergent for a given accuracy
 significantly faster.

The reader is invited to perform some additional tests by 
using several publicly available F77 codes.  
In the case of a 2D periodicity in 3D, 
numerical codes can be obtained 
from Comp. Phys. Comp.: for a complex 2D lattice in 3D see 
routines DLMNEW and DLMSET of Ref. \cite{McLC};
for a simple 2D Bravais lattice in 3D see routine XMAT 
of Ref. \cite{YSM}.
The above codes have been implemented in electronic, acoustic,
and electromagnetic LKKR codes and successfully tested time and again 
in various cases \cite{Pen,AMprb,Tong,McLC,Oh,SKM,YSM,PSM}.
A limited Windows executable which incorporates the
lattice sums within a photonic LKKR code for the calculation
of reflection, transmission, and absorption of an electromagnetic
plane  wave incident on a square array
of finite length cylinders arranged on a homogeneous slab of
finite thickness is available
following the link  http://www.wave-scattering.com/caxsrefl.exe.
 
F77 source code for a 1D Bravais periodicity in 2D is freely
available at \\ http://www.wave-scattering.com/ola.f 
(implementation
instruction are described on \newline
http://www.wave-scattering.com/dlsum1in2.html).
The code has been implemented in a corresponding  LKKR code
and successfully tested against experiment in Ref.  
\cite{PTK}. A limited Windows executable calculating
the reflection, transmission, and absorption of an electromagnetic
plane wave incident on a square array
of infinite length cylinders in the plane normal to the cylinder axis,
is available
following the link  http://www.wave-scattering.com/rta1in2k.exe.

\subsection{Outlook}
\label{sec:outl}
The present work can be straightforwardly 
extended in several directions. First, as in bulk cases \cite{Seg,HS}, 
for a 2D periodicity in 3D \cite{Kam3},
and for a 1D periodicity in 2D \cite{AmO},
the condition of a simple lattice for a 1D periodicity in 3D can 
easily be relaxed to an arbitrary periodic lattice. 
Note that the case of a non-Bravais lattice
additionally requires the calculation of the series (\ref{fsgfls})
with the origin of coordinates displaced 
from the lattice by a fixed non-zero vector. Consequently, the term 
involving $r_n=0$ is no longer singular. Therefore, in the latter case
the lattice sums are expressed as the sum of solely 
$D_{L}^{(1)}$ and $D_{L}^{(2)}$, where $D_{L}^{(2)}$ does include
the $r_n=0$ term.
These supplementary series can easily be determined 
following the recipes of Refs. \cite{Seg,HS,Bbl,Kam3}.

Second, following the work of
 Ohtaka \cite{Oh} and Modinos \cite{Mod} for a 2D periodicity in 3D,
in the vector case of electromagnetic waves for a 1D periodicity in 3D
(2D case is trivial as it reduces to a scalar problem), 
the lattice sums and structure constants 
can easily be obtained
from those in the scalar case presented here. 
It is only required
to multiply the scalar quantities with appropriate
numerical factors of geometric origin \cite{WZY,Mo,Oh,Mod}.
This possibility is a consequence of a fact, as first shown 
by Stein \cite{St}, that vector 
translational addition coefficients 
can be derived from pertinent scalar addition coefficients.
A more involved, but possible, is a generalization of the presented results
to semi-infinite cases, when periodicity is imposed on a half-line,
or in a half-space only \cite{SC,CST,Lnt1}.

\section{Summary and conclusions}
\label{sec:concl}
A classical problem of free-space Green's functions $G_{0\Lambda}$ 
representations of the Helmholtz equation was studied in various
quasi-periodic cases, i.e., when an underlying periodicity 
is imposed in less dimensions than is the dimension of an embedding space. 
Exponentially convergent series for the 
free-space quasi-periodic $G_{0\Lambda}$
and for the expansion coefficients $D_{L}$ of 
$G_{0\Lambda}$ in the basis of regular (cylindrical in 
two dimensions and spherical 
in three dimension (3D)) waves, or lattice sums,  
were reviewed and new results for the case of a one-dimensional (1D)
periodicity in 3D were derived. 
The derivation of relevant results 
highlighted the common part which is applicable to any
of quasi-periodic cases.

Exponentially convergent Ewald representations (\ref{31grfr}), 
(\ref{32grfr}), (\ref{21grfr}) 
for $G_{0\Lambda}$
 (see also Sec. \ref{sec:lplew}) and for lattice sums $D_{L}$ 
hold for any value of the Bloch momentum
and allow $G_{0\Lambda}$ to be efficiently 
evaluated also in the periodicity plane.
After substituting the resulting expressions for 
$D_{L}^{(1)}$ [Eqs. (\ref{dl1f1in3}), (\ref{dl1f2in3}), (\ref{dl1f1in2})],
$D_{L}^{(2)}$ [Eqs. (\ref{dl2f12in3}), (\ref{dl2f1in2})] and 
$D_{L}^{(3)}$ [Eqs.  (\ref{dl3f1in3}), (\ref{dl3f1in3}),
(\ref{dl3f1in2})] into defining equation (\ref{dldsum}) for 
$D_{L}$, an alternative
exponentially convergent representations for 
Schl\"{o}milch series (\ref{dstrcos}) of cylindrical and 
spherical Hankel functions of any integer order are obtained.

The quasi-periodic Green's functions of the Laplace 
equation were studied as the limiting case of
the corresponding Ewald
representations of $G_{0\Lambda}$ of the Helmholtz equation
by taking the limit
of the wave vector magnitude going to zero. 
Thereby, exponentially convergent representations of $G_{0\Lambda}$
in the Laplace case were obtained, which 
are  convergent (unless ${\bf R}\in\Lambda$) also  in the periodicity plane. 
An alternative exponentially convergent series for $G_{0\Lambda}$ 
for 1D periodicity in 3D in the Laplace case has also been obtained earlier
by Linton (see series in Eq. (3.26) of Ref. \cite{Lnt}). 
However, our expressions have been derived
without any artificial
regularization using a convergence ensuring logarithmically 
divergent series (cf. Ref. \cite{Lnt}).

The results obtained can be useful for
numerical solution of boundary integral equations 
for potential flows in fluid mechanics, 
remote sensing of periodic surfaces, periodic gratings,
in many contexts of simulating systems of charged particles,
in molecular dynamics, for solving the spectrum of 
particular open resonators,
for the description of quasi-periodic arrays of point interactions 
in quantum mechanics, linear chains of spheres and nanoparticles
in optics and electromagnetics, and of
 infinite arrays of resonators coupled to a waveguide, and in various ab-initio
first-principle multiple-scattering theories for the analysis of diffraction
of classical and quantum waves.

\newpage

\appendix

\section{Integral representations of $H_\nu^{(1)}$}
\label{app:hintr}
\begin{figure}[tbp]
\begin{center}
\epsfig{file=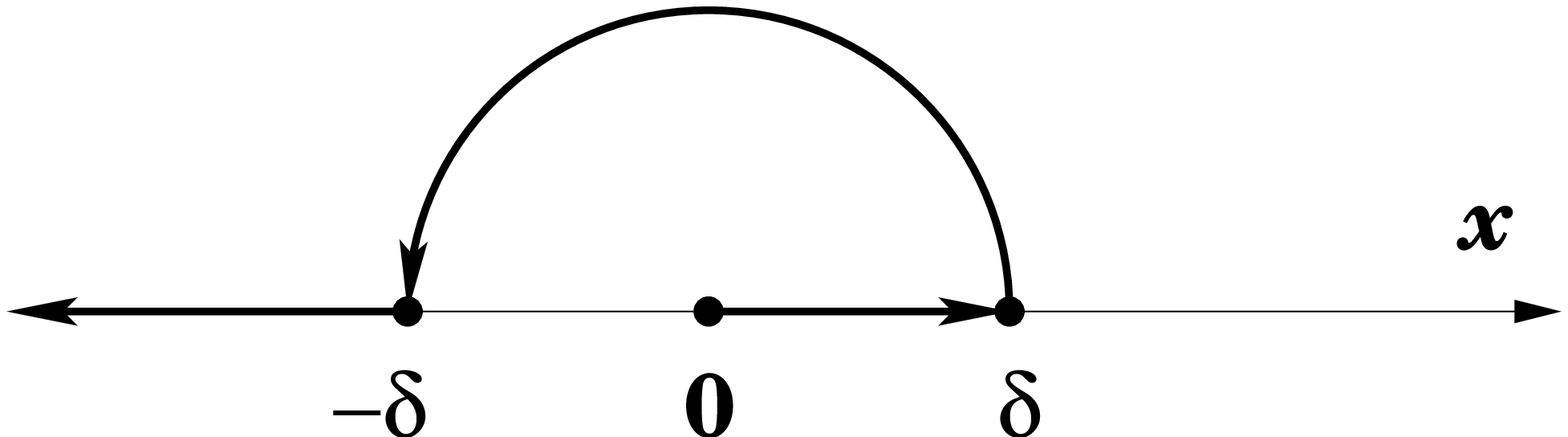,width=12cm,clip=0,angle=0}
\end{center}
\caption{}
\label{fg:schlf}
\end{figure}

\begin{figure}[tbp]
\begin{center}
\epsfig{file=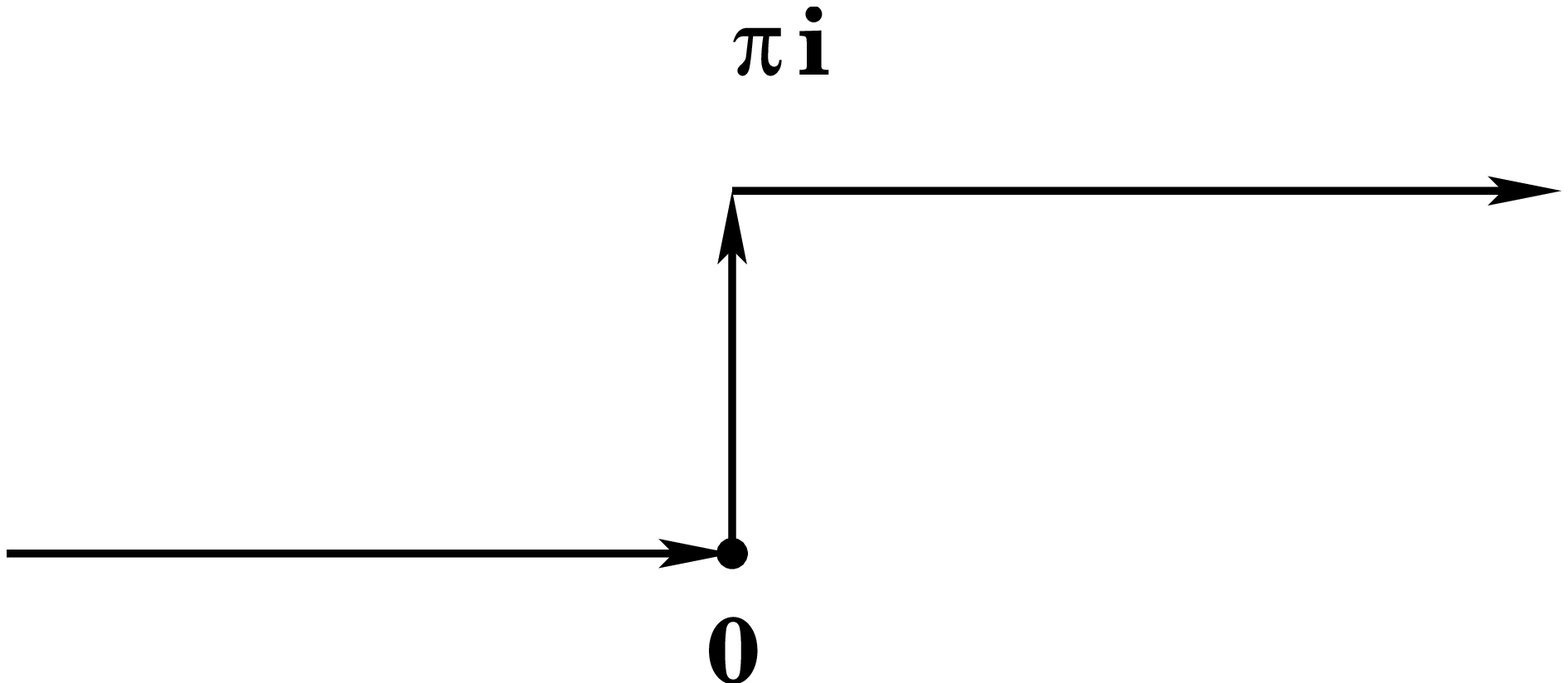,width=12cm,clip=0,angle=0}
\end{center}
\caption{}
\label{fg:intep}
\end{figure}

For the Ewald summation with ${\bf r}_s=0$ excluded,
one uses the Schl\"{a}fli integral representation,
\begin{equation}
H_\nu^{(1)}(z)=\frac{1}{\pi i}\int_{C_-}
u^{-\nu-1} e^{\frac{1}{2}\,z\left(u-\frac{1}{u}\right) }\, du,
\label{schlaf}
\end{equation}    
where the contour $C_-$ is the contour which goes from the origin
to $\delta>0$, continues along a semicircle in the upper
half-plane with radius $\delta$ to $-\delta$, and  goes along
the negative real axis to infinity (see Fig. \ref{fg:schlf}). This results 
in an exponentially decreasing integrand at the 
integration contour ends for Re $z>0$. 
The Schl\"{a}fli  representation, which yields the Hankel functions
$H_\nu^{(1)}(z)$ as moments of the generating function of the Bessel 
functions $J_l(z)$ (see \cite{Wat}, p. 14), is obtained upon 
substitution $u=e^t$ in the integral representation 
(see (9.1.25) of Ref. \cite{AS}),  
\begin{equation}
H_\nu^{(1)}(z)=\frac{1}{\pi i}\int_{-\infty}^{\infty+\pi i}
e^{z\sinh t -\nu t}\, dt,
\label{asr}
\end{equation}
taken along the contour shown in Fig. \ref{fg:intep},
which is valid for $|\arg z|<\pi/2$.

\begin{figure}[tbp]
\begin{center}
\epsfig{file=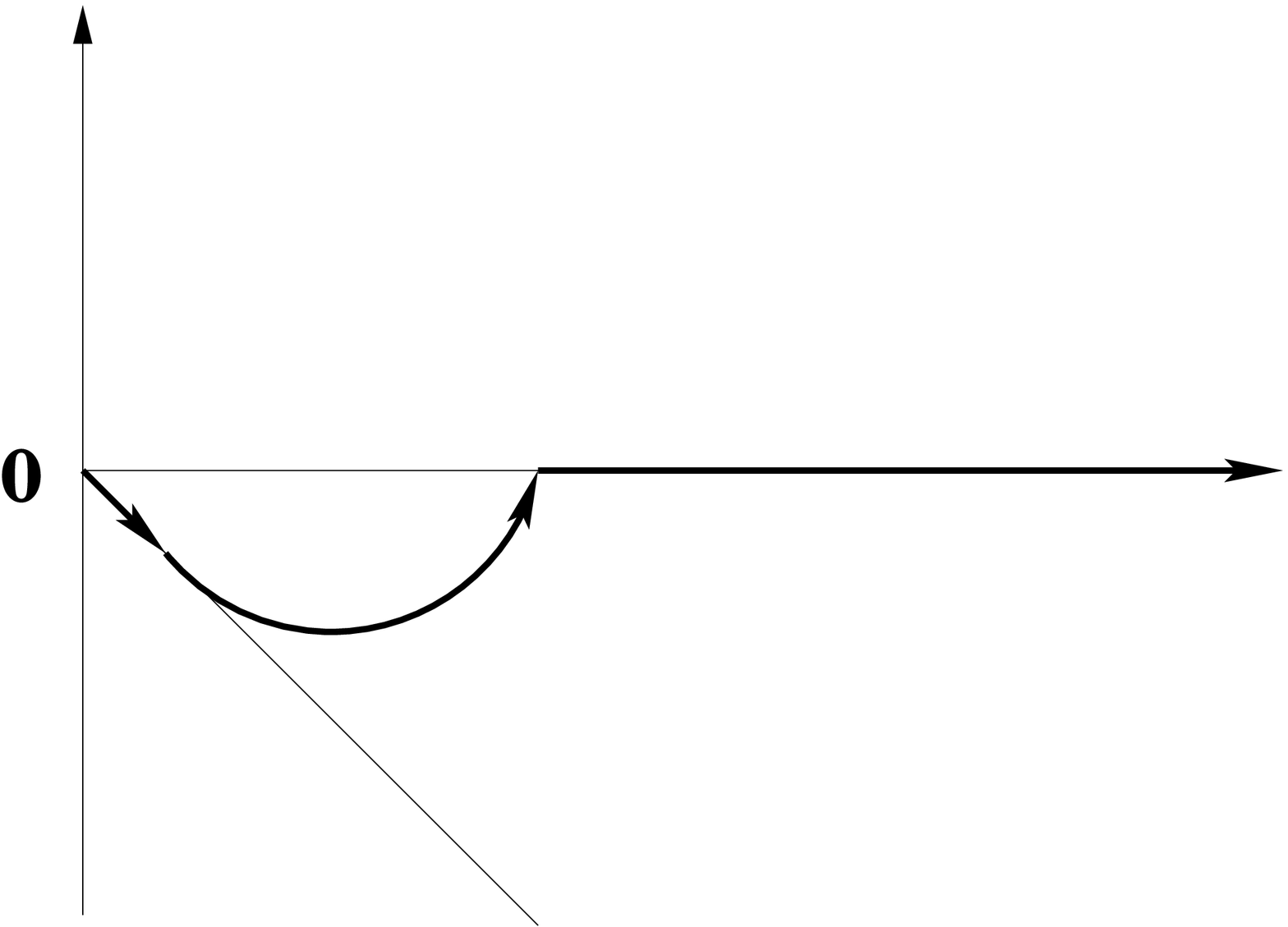,width=12cm,clip=0,angle=0}
\end{center}
\caption{}
\label{fg:ewald}
\end{figure}

Setting in the Schl\"{a}fli representation (\ref{schlaf}) $z=\sigma r$ and 
\begin{equation}
t=-\frac{2r}{\sigma}\,\xi^2,
\hspace*{1.2cm} du=-2 \left(\frac{\sigma}{2r}\right)^{-1}\,\xi d\xi,
\end{equation}
one arrives at
\begin{equation}
H_{\nu}^{(1)}(\sigma r)=\frac{2 i^{-2\nu-1}}{\pi} 
                       \left(\frac{\sigma}{2r}\right)^\nu 
\int_{C}
\xi^{-2\nu-1}  e^{- r^2\xi^2+\sigma^2/(4\xi^2)} \, d\xi.
\label{xiint}
\end{equation}
where $C$ is the so-called Ewald contour (see Fig. \ref{fg:ewald}),
which leaves the origin 
along the ray arg $\xi=\mbox{arg }\sigma -\pi/4$, then returns to the real 
axis and continues 
along the positive real axis to infinity.

Upon substituting in the Schl\"{a}fli representation (\ref{schlaf})  
$z=\sigma r$ and 
\begin{equation}
u =\frac{r}{\sigma}\,\zeta,
\label{resu}
\end{equation}
one arrives at
\begin{equation}
H_\nu^{(1)}(\sigma r)=\frac{1}{\pi i} \left(\frac{\sigma}{r}\right)^\nu  \int_{C}
\zeta^{-\nu-1} e^{\frac{1}{2}\, (r^2\zeta - \sigma^2/\zeta)}\, d\zeta.
\label{kmb32}
\end{equation} 
Let us consider for a while $\sigma$ as a general real parameter. 
Then, following discussion at the end of Appendix 2 of Ref. \cite{Kam1}, 
the Schl\"{a}fli integration contour for a positive
$\sigma^2$ can be deformed to that shown in Fig. \ref{fg:kmb}. 
Afterward, with the use 
of Jordan's lemma (\cite{WWat}, p. 115) integration contour can be deformed 
to that from $0$ to $i\infty$ on the imaginary axis.
For a negative $\sigma^2$ one would then, as the result of
the substitution (\ref{resu}), arrive at the contour from $0$ to $\infty$ 
along the positive real axis (see Appendix 2 of Ref. \cite{Kam1}).

Eventually we provide Hobson's representation for the modified
Bessel functions of the third kind \cite{GlZ}
\begin{equation}
\left(\frac{q}{k}\right)^\nu K_\nu(kq)=     \int_0^\infty
\zeta^{\nu-1} e^{-\frac{1}{2}\, (k^2\zeta +q^2/\zeta)}\, d\zeta.
\label{hobs}
\end{equation}

\section{Properties of harmonics ${\cal Y}_L$} 
\label{app:sphr}
Throughout this paper complex harmonics
${\cal Y}_L$ (cylindrical, $Y_l=e^{il\phi}/\sqrt{2\pi}$, for $d=2$ and spherical 
for $d=3$ \cite{AS,Kam3,Pen,Tong}; for 1D harmonics see Ref. \cite{But})
are used.
Under complex conjugation, they behave according to 
\begin{equation}
{\cal Y}_L^*(\hat{{\bf R}})=\left\{
\begin{array}{cc} 
{\cal Y}_{l}(\hat{{\bf R}}),  &  1D\\ 
{\cal Y}_{-l}(\hat{{\bf R}}), &2D\\
(-1)^m {\cal Y}_{l-m}(\hat{{\bf R}}), & 3D.
\end{array}\right.
\label{ylcc}
\end{equation} 
In  3D the property goes under the name of
 the Condon-Shortley convention.
In any dimension, the harmonics  satisfy the inversion formula 
\begin{equation}  
{\cal Y}_{L}(-\hat{{\bf r}}) = (-1)^l {\cal Y}_{L}(\hat{{\bf r}}),
 \label{rinv} 
\end{equation} 
the orthonormality 
\begin{equation}
\oint {\cal Y}_{L}(\hat{{\bf r}}){\cal Y}_{L'}^*(\hat{{\bf r}})\, d\Omega =\delta_{LL'},
\label{shorth}
\end{equation} 
and closure 
\begin{equation}
 \sum_{L}  {\cal Y}_{L}(\hat{{\bf r}}){\cal Y}_{L}^*(\hat{{\bf r}}')
=\delta_\Omega({\bf r},{\bf r}'). 
\nonumber
\end{equation} 
Here 
$\delta_\Omega({\bf r},{\bf r}')$ is the delta function on the unit sphere
whereas 
$d\Omega$ is the usual angular measure which is determined 
by the relation $d{\bf r} =r^{d-1}dr d\Omega$. 
(In 1D the angular integral $\oint d\Omega$ 
reduces to the summation over the forward and backward directions.)

Note that in any dimension,
\begin{equation}
{\cal Y}_0=\frac{1}{\sqrt{A}}\cdot 
\label{yps0} 
\end{equation}
Since ${\cal Y}_0$ is a constant, combining
Eq. (\ref{yps0}) with the orthonormality (\ref{shorth}) of ${\cal Y}_{L}$
yields
\begin{equation}
\oint {\cal Y}_{L}(\hat{{\bf r}})\, 
   d\Omega = \sqrt{A} \, \delta_{L0}.
\label{sphryps0}
\end{equation}

\subsection{The plane-wave expansion}
An expansion of the plane-wave expansion in 
the angular momentum basis is  
\begin{equation} 
e^{i{\bf k}\cdot {\bf r}}= A \sum_{L} i^{|l|} {\cal J}_{|l|}(kr)\,
 {\cal Y}_L(\hat{{\bf r}}) {\cal Y}_L^*(\hat{{\bf k}})= 
A \sum_{L} i^{|l|} {\cal J}_{|l|}(kr)\, 
{\cal Y}_L^*(\hat{{\bf r}}) {\cal Y}_L(\hat{{\bf k}}), 
\label{gplw} 
\end{equation} 
where 
\begin{equation} 
A=\oint d\Omega = \left\{ 
\begin{array}{cc}
 2, & 1D\\ 
2\pi, & 2D\\ 
4\pi, & 3D 
\end{array}\right.
\label{Adef} 
\end{equation} 
where $d\Omega$ here is the usual angular measure. 
The series (\ref{gplw}) converges uniformly as $|{\bf k}|$ and ${\bf r}$
run through compact sets of $\mathbb{R}$ and $\mathbb{R}^3$ (Theorem XI.64f
of Ref. \cite{RS3}).
The absolute value of $l$ is used in (\ref{gplw})
in case  the sum over angular momenta in 2D runs 
from minus to plus infinity.

The plane-wave expansion (\ref{gplw}) is also valid for 
complex arguments. 
It is interesting to note that ${\cal Y}_{L}^*$ is no longer
the complex conjugate of ${\cal Y}_{L}$ for complex ${\bf k}$
(e.g., in 3D because of the complex
nature of associated Legendre functions). However, the relations (\ref{ylcc})
remain the same as in the case of harmonics 
of a real argument \cite{Kam3,Tong}. 
For 3D case see, for instance, Appendix 1 of Ref. \cite{Kam3}
or Appendix A of review by Tong \cite{Tong}.
The 2D case follows by a straightforward adaptation of the 3D case, whereas
the 1D case is trivial \cite{But}.

\section{Free-space scattering Green's function}
\label{app:fsgf}  
One has [see Eq.(\ref{fsgf})]
\begin{equation} 
G_0^+ (\sigma,{\bf r},{\bf r}') = \frac{1}{(2\pi)^d}
 \int \frac{e^{i{\bf k}\cdot({\bf r}-{\bf r}')}}{\sigma^2-k^2+ i\epsilon}\,d^d{\bf k} 
= -i C {\cal H}_0^+(\sigma |{\bf r}-{\bf r}'|), 
\nonumber
\end{equation} 
where  
\begin{equation} 
C= \frac{\pi}{2}\, \frac{A}{(2\pi)^d}\, \sigma^{d-2} 
=\left\{ 
\begin{array}{ll} 
\frac{1}{2\sigma},   &   1D\\ 
\frac{1}{4},      &   2D\\ 
\frac{\sigma}{4\pi}, & 3D 
\end{array} \right. 
\label{CArel}
\end{equation} 
 is a real positive number for positive energies. 
The 2nd equality in (\ref{fsgf}) is established by expanding the exponential
into regular waves according to Eq. (\ref{gplw}), performing
the angular integral using
the identity (\ref{sphryps0}) satisfied by the harmonics ${\cal Y}_L$, 
and eventually performing the remaining 
radial integral using the integral identity 
\begin{equation}
I=\int_0^\infty 
\frac{{\cal J}_0(kr)}{\sigma^2+i\epsilon-k^2}\, k^{d-1}
dk = -\frac{\pi i}{2}\, \sigma^{d-2} {\cal H}^{+}_0(\sigma r),
\label{int0}  
\end{equation}
which can be easily established by contour integration in complex plane.
In arriving at Eq. (\ref{int0}) one substitutes for $J_0$ according to 
\begin{equation}
{\cal J}_l(kr)=\frac{1}{2}\left[
{\cal H}_l^{+}(kr)+{\cal H}_l^{-}(kr)\right].
\label{jltohl}
\end{equation}
and  applies the identity (analyticity property) 
\begin{equation}
{\cal H}_l^-(ze^{-\pi i})= (-1)^{l+d+1} {\cal H}_l^+(z)
\label{reflh-} 
\end{equation}
(see Eqs. (9.1.16) and (10.1.18) of Ref. \cite{AS} and Ref. \cite{But}) for $l=0$.

 One can show that 
\begin{equation}
 |{\bf x} -{\bf y}|\sim |{\bf x}| - \frac{{\bf x}\cdot{\bf y}}{|{\bf x}|} + 
{\cal O} \left(1/|{\bf x}|^2\right)\hspace*{1.5cm} \mbox{as } |{\bf x}|\rightarrow \infty. 
\nonumber
\end{equation} 
Therefore, upon using the asymptotic properties (\ref{hlas}) of 
${\cal H}_0^+ (\sigma |{\bf x}-{\bf y}|)$ 
for $|{\bf x}|\rightarrow \infty$, 
\begin{equation} 
G_0^+ ({\bf x},{\bf y}) \sim f_\sigma(|{\bf x}|) e^{-i\sigma {\bf x}\cdot{\bf y}/|{\bf x}|} 
= f_\sigma(|{\bf x}|) e^{-i{\bf k}'\cdot{\bf y}}, 
\label{fsgfas} 
\end{equation} 
where 
\begin{equation} 
f_\sigma(|{\bf x}|)\sim G_0^+({\bf x},{\bf 0})|_{{\bf x}\rightarrow\infty} 
= -i C {\cal H}_0^+  (\sigma|{\bf x}|),\hspace*{1.5cm} {\bf k}'
 = \sigma{\bf x}/|{\bf x}|. 
\label{fsgvh0} 
\end{equation} 
Function $f_\sigma(|{\bf x}|)$ describes outgoing waves in a 
given dimension. Explicitly 
\begin{equation} 
f_\sigma(|{\bf x}|)  
=\left\{ 
\begin{array}{ll} 
- \frac{i}{2\sigma} \, e^{i\sigma|{\bf x}|}, & 1D\\ 
-\frac{i}{\sqrt{8\pi\sigma}} 
\frac{e^{i\sigma|{\bf x}|- i\pi/4} } {\sqrt{|{\bf x}|}}, & 2D\\ 
-\frac{1}{4\pi} \frac{e^{i\sigma|{\bf x}|}}{|{\bf x}|}, & 3D.
 \end{array}\right. 
\label{fsgas} 
\end{equation}

The product AC in the partial wave expansion (\ref{gfpwex}) 
of the free Green's 
function can be independently determined 
 from the condition 
that the discontinuity  of  radial 
derivatives of the free Green's function at
 coinciding arguments multiplied by $r^{d-1}$ 
is exactly one. The factor  $r^{d-1}$ follows 
from a fact that the integral measure $d{\bf r}$ can be 
written as $d{\bf r} = r^{d-1} dr \, d\Omega$. 
Since the harmonics ${\cal Y}_L$ are 
orthonormal in the measure $d\Omega$ [Eq. (\ref{shorth})], 
the discontinuity  of radial 
derivatives of the free Green's function can be conveniently
expressed by the Wronskian  
\begin{equation} 
W_x[f(ax),g(ax)]=f(ax) g'(ax)-f'(ax)g(ax),
\nonumber
\end{equation} 
where prime denotes first derivative with respect to $x$.  
Now
$$W_r[{\cal J}_l(\sigma r),{\cal N}_l(\sigma r)] 
=\sigma W_z[{\cal J}_l(z),{\cal N}_l(z)]$$
for 
$z=\sigma r$ and 
$$W_r[{\cal J}_l(\sigma r),{\cal H}_l^+(\sigma r)] 
= i W_r[{\cal J}_l(\sigma r),{\cal N}_l(\sigma r)].$$ 
Knowing the Wronskian  
\begin{equation} 
W_z[{\cal J}_l(z),{\cal N}_l(z)] 
 ={\cal J}_l {\cal N}_l'- {\cal J}_l'{\cal N}_l 
 =\left\{  
\begin{array}{cl} 
1, & 1D\\ 
2/(\pi z), & 2D \\ 
1/z^2, & 3D, 
\end{array}\right. 
\nonumber
\end{equation} 
the product AC can be then found directly from
the discontinuity given by the relation 
\begin{equation} 
AC = \left\{\sigma r^{d-1} 
W_z\left[{\cal J}_l(z),{\cal N}_l(z)\right] \right\}^{-1}.
\nonumber 
\end{equation}

\section{Jacobi identities}
\label{app:jacobi}
The equivalence of the lattice sum 
[see Eq. (\ref{lgrfdf})] and the eigenvalue expansion [see Eq. (\ref{dualbulk})] 
in the case of the heat equation with the Bloch boundary conditions
in a box leads to the identity
\begin{equation}
K_\Lambda ({\bf R},t)=\frac{1}{(4\pi t)^{d/2}}\, 
  \sum_{{\bf r}_s\in\Lambda}
     e^{i{\bf k}\cdot{\bf r}_s} e^{-({\bf R}-{\bf r}_s)^2/(4t)} =\frac{1}{v_0} 
\sum_{{\bf k}_n\in\Lambda^*} 
  e^{-({\bf k}+{\bf k}_n)^2 t}e^{i({\bf k}+{\bf k}_n)\cdot{\bf x}}.
\label{dide}
\end{equation}
In a special case, for a simple  cubic lattice with a unit 
lattice constant one has $\Lambda=\Lambda^*$. Upon substituting
${\bf k}_n=2\pi {\bf n}$, ${\bf n}$ being an integer valued vector, and
${\bf R}={\bf k}=0$, identity (\ref{dide}) yields 
\begin{equation}
\frac{1}{(4\pi t)^{d/2}}\, \sum_{{\bf n}\in\Lambda}
e^{-{\bf n}^2/(4t)} = \sum_{{\bf n}\in\Lambda} 
e^{-4\pi^2 {\bf n}^2 t}.
\nonumber
\end{equation}
For $d=1$, the latter is a special  $\theta=0$ case
of the famous number theoretical Jacobi theta function identity
(upon rescaling $t\rightarrow 4\pi t$),
\begin{equation}
\sum_{n=-\infty}^\infty \exp(-\pi n^2 t-2\pi i n\theta) =
t^{-1/2}
\sum_{l=-\infty}^\infty \exp\left[-\pi (l +\theta)^2/t\right],
\label{thetaid} 
\end{equation}
which is valid for a complex $\theta$ and Re $t>0$.  
The Jacobi theta function identity can also be proved
by applying the Poisson sum rule and is also sometimes referred to as
Poisson-Jacobi formula.

An alternative form (\ref{edikmb}) of the Jacobi formula,  
as has been used by Kambe \cite{Kam3,Kam2,Kam1},
is obtained upon substitution $t=\zeta/2$ as is valid for
Re $\zeta>0$.
After the substitution $t=1/(4\xi^2)$, the 
generalized Jacobi identity (\ref{dide}) yields 
\begin{equation}
\sum_{{\bf r}_s\in\Lambda} e^{-({\bf R}-{\bf r}_s)^2\xi^2 
+i{\bf k}\cdot{\bf r}_s}=
\frac{\pi^{d/2}}{v_0 \xi^d}\sum_{{\bf k}_n\in\Lambda^*}
e^{-({\bf k}_n+{\bf k})^2/(4\xi^2) +i({\bf k}+{\bf k}_n)\cdot{\bf R}}.
\label{edi}
\end{equation} 
In like manner, upon the substitution ${\bf R}\rightarrow -{\bf R}$, 
${\bf k} \rightarrow -{\bf k}$, 
${\bf k}_n \rightarrow -{\bf k}_n$,  one finds
\begin{equation}
\sum_{{\bf r}_s\in\Lambda} e^{-({\bf R}+{\bf r}_s)^2\xi^2 
- i{\bf k}\cdot({\bf x}+{\bf r}_s)}=
\frac{\pi^{d/2}}{v_0 \xi^d}\sum_{{\bf k}_n\in\Lambda^*}
e^{-({\bf k}_n+{\bf k})^2/(4\xi^2) + i{\bf k}_n\cdot{\bf R}}.
\label{edileed}
\end{equation} 
The last two identities are often referred to as the Ewald identities 
(cf. Refs. \cite{HS,Ew1,Ew2}).

\section{McRae derivation of  dual representations}
\label{app:fsqpgf}  
In the quasi-periodic case, the dual representation 
can also be established by applying 
an Ewald integral representation of Green's function
and the generalized Jacobi identity.
This path has been originally followed by McRae \cite{McRaer} 
for a 2D periodicity in 3D.
Here it will also be outlined for the remaining two cases.

Let 
\begin{equation}
S_{d,d_\Lambda} (\sigma,{\bf r}) = \sum_{{\bf r}_s\in\Lambda} 
{\cal H}_0^+(\sigma|{\bf r}+{\bf r}_s|)
e^{-i{\bf k}\cdot({\bf r}_s+{\bf r})},
\label{ddl}
\end{equation}
where $\Lambda$ is as usual a $d_\Lambda$-dimensional lattice ($d_\Lambda<d$) and
${\bf r}_s$ are the lattice points. 
Then [cf. Eq. (\ref{fsgf})]
\begin{equation} 
G_{0\Lambda} (\sigma,{\bf r},{\bf r}') = 
 -i \frac{\pi}{2}\, \frac{A}{(2\pi)^d}\, \sigma^{d-2} 
S_{d,d_\Lambda} (\sigma,{\bf r}-{\bf r}'). 
\label{lgfvls} 
\end{equation} 

Let us first consider a special case
\begin{equation}
S_{3,d_\Lambda} = \sum_{{\bf r}_s\in\Lambda} h_{0}^{(1)}(\sigma|{\bf r}+{\bf r}_s|)
e^{-i{\bf k}\cdot({\bf r}_s+{\bf r})}.
\label{31d}
\end{equation}
Now the Ewald integral representation 
[Eq. (\ref{xiint}) of Appendix \ref{app:hintr}] 
\begin{equation}
h_{0}^{(1)}(\sigma r)= -\frac{2i}{\sqrt{\pi}\sigma}                     
\int_{C}
 e^{\sigma^2/(4\xi^2)- r^2\xi^2} \, d\xi
\label{rixiint}
\end{equation}
is substituted for $h_{0}^{(1)}$ in Eq. (\ref{31d}).
Here the path of integration leaves the origin along the ray
arg $\xi=\mbox{arg }\sigma -\pi/4$ and then returns to the real axis.
Afterward one finds
\begin{equation}
S_{3,d_\Lambda} = -\frac{2i}{\sqrt{\pi}\sigma} \sum_{{\bf r}_s\in\Lambda}                     
\int_{C}
 e^{\sigma^2/(4\xi^2)- [(r_s+{\bf r}_\parallel)^2 +r_\perp^2]\xi^2} 
e^{-i{\bf k}\cdot({\bf r}_s+{\bf r})}\, d\xi.
\end{equation}
By regarding the integral as the limit Im $\sigma\rightarrow 0_+$,
the double summation here can be shown to be 
absolutely convergent for $r_\perp\neq 0$,
and hence the order of contour integration and summation can be reversed.
Indeed, on the interval $(\eta,\infty)$, $\eta>0$, along the positive real axis
the sum of integrands is less than
\begin{equation}
  \mbox{const}\times e^{-r_\perp^2 \xi^2}\cdot
\nonumber
\end{equation}
On the other hand, in the proximity of zero along the Ewald integration contour 
$\xi= e^{-i\pi/4} t$, where $t\in (0,\eta)$ is a real positive number, and hence
the sum of integrands is less than
\begin{equation}
\frac{\mbox{const} }{1-e^{it^2}} \times e^{-\sigma''/t^2},
\nonumber
\end{equation}
where $\sigma''$ denotes the imaginary part of $\sigma$.
After the reversal of summation and integration, the respective 
Jacobi identities (\ref{edileed}) for $d_\Lambda=1$ 
and $d_\Lambda=2$ can be applied.
To this end, the respective expressions for 
$d_\Lambda=1$ and $d_\Lambda=2$
are formally identical and they only differ in the dimensionality 
of the lattice $\Lambda$. 

Upon using the generalized Jacobi identity (\ref{edileed})
for $d_\Lambda=1$ with ${\bf R}={\bf r}_\parallel$ one obtains
\begin{equation}
S_{3,1} =-\frac{2i}{v_0\sigma} 
 \sum_{{\bf k}_n\in\Lambda^*} 
e^{i{\bf k}_n\cdot{\bf r}_\parallel}        
\int_{C}
 e^{K_{\perp n}^2/(4\xi^2)- r_\perp^2\xi^2}\, \xi^{-1} d\xi,
\end{equation}
where $K_{\perp n}$ is as usual, given by Eq. (\ref{Kkn}), and
$\Lambda^*$ is  a corresponding reciprocal 1D lattice with 
lattice points ${\bf k}_n$.
Now, for $r>0$ and $|\arg \sigma|<\pi/2$, Eq. (\ref{xiint}) yields
\begin{equation}
H_{0}^{(1)}(\sigma r)=\frac{2}{\pi i} \int_{C}
 e^{\sigma^2/(4\xi^2)- r^2\xi^2} \, \xi^{-1} d\xi.
\label{H0int}
\end{equation}
Therefore,
\begin{equation}
S_{3,1} = \frac{\pi}{v_0\sigma}  \sum_{{\bf k}_n\in\Lambda^*} 
e^{i{\bf k}_n\cdot{\bf r}_\parallel} 
  H_{0}^{(1)}(K_{\perp n} |r_\perp|).
\label{31dualo}
\end{equation}

For $d_\Lambda=2$, 
upon using the generalized Jacobi identity (\ref{edileed})
 with ${\bf x}={\bf r}_\parallel$, one obtains
\begin{equation}
S_{3,2}=\frac{-i \pi}{v_0 \sigma} \sum_{{\bf k}_n\in\Lambda^*} 
e^{i{\bf k}_n\cdot{\bf x}} \,
\frac{2}{\sqrt{\pi}} \int_{C} \sum_{{\bf k}_n\in\Lambda^*} 
e^{K_{\perp n}^2/(4\xi^2)-r_\perp^2\xi^2 }\, \xi^{-2} d\xi.
\label{32dualp}
\end{equation}
Upon substituting 
\begin{equation}
\xi^{-2}=\int_0^\infty e^{-\xi^2 t}\, dt,
\label{xi2tric}
\end{equation}
which is a special case of the Euler transformation \cite{Ep,GlZ}, 
interchanging the order of integration and using 
[Eq. (\ref{rixiint}); Appendix \ref{app:hintr}]
\begin{equation}
\frac{e^{i\sigma r}}{r}=  \frac{2}{\sqrt{\pi}}\int_{C}
 e^{\sigma^2/(4\xi^2) - r^2 \xi^2}\, d\xi,
\label{ewin}
\end{equation}
the integral on the r.h.s. of Eq. (\ref{32dualp}) becomes
\begin{equation}
\frac{2}{\sqrt{\pi}} \int_{C} 
e^{K_{\perp n}^2/(4\xi^2)-r_\perp^2\xi^2 }\, \xi^{-2} d\xi=
\int_0^\infty 
\frac{e^{iK_{\perp n}(r_\perp^2+t)^{1/2}}}{(r_\perp^2+t)^{1/2}}
\, dt. 
\label{sintp}
\end{equation}
Upon substituting 
\begin{equation}
u=(r_\perp^2+t)^{1/2}, \hspace*{1.4cm} 
du=dt/[2(r_\perp^2+t)^{1/2}],
\label{sfort}
\end{equation}
one eventually arrives at
\begin{equation}
\int_0^\infty 
\frac{e^{iK_{\perp n}(r_\perp^2+t)^{1/2}}}{(r_\perp^2+t)^{1/2}}
\, dt=
2 \int_{r_\perp}^\infty e^{i K_{\perp n}u}\, du=
2i \frac{e^{i K_{\perp n}|r_\perp|}}{K_{\perp n}}
= -2 |r_\perp| h_{0}^{(1)}(K_{\perp n}|r_\perp|).  
\label{sint}
\end{equation}
For real $K_{\perp n}$ one adopts as the value of the 
integral the limit as Im $K_{\perp n}\rightarrow 0_+$.
Eventually,
\begin{equation}
S_{3,2} = \frac{2 \pi i}{v_0\sigma}  \sum_{{\bf k}_n\in\Lambda^*} 
e^{i{\bf k}_n\cdot{\bf r}_\parallel} 
  |r_\perp| h_{0}^{(1)}(K_{\perp n}|r_\perp|).
\label{32dualo}
\end{equation}

For completeness, and  without a proof, we  also present the result
for
\begin{equation}
S_{2,1} = \sum_{{\bf r}_s\in\Lambda} H_{0}^{(1)}(\sigma|{\bf r}+{\bf r}_s|)
e^{-i{\bf k} \cdot({\bf r}_s+{\bf r})} = \frac{2i}{v_0} 
\sum_{{\bf k}_n\in\Lambda^*} e^{i{\bf k}_n\cdot{\bf r}_\parallel}\,
 |r_\perp| h_{0}^{(1)}(K_{\perp n}|r_\perp|),
\label{21dualo}
\end{equation}
which have been
used as an intermediary step in the derivation of results
of Ref. \cite{Mol}.  

When the respective dual representations (\ref{31dualo}), (\ref{32dualo}), 
(\ref{21dualo}) 
are substituted into Eq. (\ref{lgfvls}) for 
a corresponding $G_{0\Lambda}$, ${\bf r}$ in 
Eqs. (\ref{31dualo}), (\ref{32dualo}), 
(\ref{21dualo}) is substituted by ${\bf R}$.

\section{General properties of free-space quasi-periodic  
Greens functions and of lattice sums} 
\label{sec:gfgenprlc}
For any ${\bf r}_s\in\Lambda$, ${\bf k}_n\in\Lambda^*$,    
$G_{0\Lambda}$ satisfies the following trivial properties,  
\begin{equation}
G_{0\Lambda}(\sigma,{\bf k}_\parallel,{\bf R}) 
=G_{0\Lambda}(\sigma,{\bf k}_\parallel,{\bf R}+{\bf r}_s)
=G_{0\Lambda}(\sigma,{\bf k}_\parallel+{\bf k}_n,{\bf R}). 
\nonumber
\end{equation} 
Obviously, $G_{0\Lambda}$ is only a function of the projection 
${\bf k}_\parallel$ of ${\bf k}$ upon $\Lambda^*$ 
(note that ${\bf k}_\parallel={\bf k}$ for $d_\Lambda=d$).

Except for the frequencies which satisfy 
$\sigma^2=({\bf k}_\parallel+{\bf k}_n)^2$ 
for some ${\bf k}_n\in\Lambda^*$, $G_{0\Lambda}$ 
satisfies the following {\em reflection 
symmetry} property \cite{Kam2,Kam1},   
\begin{equation} 
G_{0\Lambda}(\sigma,{\bf k}_\parallel,{\bf R}_\parallel+{\bf R}_\perp)                
=G_{0\Lambda}(\sigma,{\bf k}_\parallel,{\bf R}_\parallel-{\bf R}_\perp). 
\label{ordpr}
\end{equation}     
From dual representations (\ref{31dual}), 
(\ref{32dual}) it follows that the respective
$G_{0\Lambda}$ are {\em hermitian} for the interchange of variables 
${\bf r}_\parallel$ and ${\bf r}_\parallel'$
and 
{\em complex symmetric} for the interchange of variables 
$r_\perp$ and $r_\perp'$.

Obviously, for any ${\bf r}_n\in\Lambda$ also $-{\bf r}_n\in\Lambda$.
Upon combing the inversion formula (\ref{rinv}) of angular-momentum harmonics
${\cal Y}_L$ with the defining Eq. (\ref{dstrcos}) for $D_{L}$ 
one finds
\begin{equation}
D_{L}(\sigma,{\bf k}_\parallel)
= -iCA^{1/2}\delta_{L0} - i AC
\sum_{{\bf r}_n\in\bar{\Lambda}}{}^{'}  {\cal H}_{l}^+(\sigma r_n) 
{\cal Y}_L^* (\hat{{\bf r}}_n) \left\{
\begin{array}{cc}
 2 \cos ({\bf k}\cdot{\bf r}_n), & l \mbox{ even}\\
 2i \sin ({\bf k}\cdot{\bf r}_n), & l \mbox{ odd}.
\end{array}\right.
\label{dstrcs} 
\end{equation}
The summation here is performed over 
the subset $\bar{\Lambda}\subset \Lambda$ of equivalence
classes of ${\bf r}_n\in\Lambda$ with respect to spatial inversion
${\bf r}_n\rightarrow -{\bf r}_n$.
Thus, for $l$ {\em even} the corresponding lattice sums
 $D_{L}(\sigma,{\bf k}_\parallel)$ are {\em even} functions
of the Bloch vector ${\bf k}_\parallel$,
\begin{equation}
D_{L}(\sigma,{\bf k}_\parallel) = D_{L}(\sigma,- {\bf k}_\parallel).
\nonumber
\end{equation}
In the special case of a 2D periodicity in 3D one has
additionally
\begin{equation}
D_{L}(\sigma,{\bf k}_\parallel) 
= D_{L}(\sigma,{\bf k}_\parallel^\pm) = 
D_{L}(\sigma,{\bf k}_\parallel^\mp ),
\nonumber
\end{equation}
where the respective wave vectors ${\bf k}^\pm_\parallel =(k_1,-k_2)$, 
 ${\bf k}_\parallel^\mp =(-k_1,k_2)$ are formed
from the Bloch vector ${\bf k}_\parallel$ components $k_1$ and $k_2$.

In the special case of $D_{00}(\sigma,{\bf k}_\parallel)$ 
for $\sigma=i\sqrt{-z}$ there are known some further 
general analytic properties 
in the complex $z$-plane, which have been
established by Karpeshina \cite{Kar,Krp}: 

\begin{itemize}

\item $D_{00}(i\sqrt{-z},{\bf k}_\parallel)$ is for a fixed Bloch 
vector ${\bf k}_\parallel$ an analytic function on
the complex plane cut along the real half-axis $z> |{\bf k}_\parallel|^2$ 
with Im $D_{00}\neq 0$ on both sides of the cut. 

\item 
On the real half-axis $z < |{\bf k}_\parallel|^2$ the function 
$D_{00}(i\sqrt{-z},{\bf k}_\parallel)$ is real, smooth, 
and increases monotonically 
($0<\partial_z D_{00}(i\sqrt{-z},{\bf k}_\parallel) <\infty$) 
from $-\infty$ to $\infty$.

\end{itemize}

\section{Alternative definitions of lattice sums and structure constants}
\label{app:alter}
If instead of $D_\Lambda$ of Eq. (\ref{dlmbd}) the difference 
\begin{equation}
{\cal D}_\Lambda(\sigma,{\bf k}_\parallel,{\bf R})
 = G_{0\Lambda}(\sigma,{\bf k},{\bf R}) - G_0(\sigma,{\bf R}),
\label{dlmbdm}
\end{equation}
which 
is also regular for ${\bf R} \rightarrow 0$,  
is expanded in terms of the  
regular  waves, 
\begin{eqnarray}  
\lefteqn{  
{\cal D}_\Lambda (\sigma,{\bf k}_\parallel,{\bf R}) 
=\sum_{L} {\cal D}_{L}(\sigma,{\bf k}_\parallel)  
{\cal J}_{l}(\sigma R) {\cal Y}_L(\hat{{\bf R}}) } \label{dstrc}
\\ 
 && =  
\sum_{L,L'} g_{LL'}(\sigma,{\bf k}_\parallel) {\cal J}_{l}(\sigma r)  
{\cal J}_{l'}(\sigma r') {\cal Y}_L(\hat{{\bf r}}){\cal Y}_{L'}^*(\hat{{\bf r}}').  
\label{dstrcoh}  
\end{eqnarray}  
this results in alternative 
lattice sums ${\cal D}_{L}$ and structure constants 
$g_{LL'}$. 
We recall here that the structure constants can in a known 
way be unambiguously 
determined from the lattice sums \cite{KR}.

The lattice sums ${\cal D}_{L}$ and $D_{L}$ 
and structure constants  $A_{LL'}$ and $g_{LL'}$,
where
\begin{equation}  
D_\Lambda(\sigma,{\bf k}_\parallel,{\bf R}) 
=  \sum_{L,L'} A_{LL'}(\sigma,{\bf k}_\parallel) {\cal J}_{l}(\sigma r)  
 {\cal J}_{l'}(\sigma r') {\cal Y}_L(\hat{{\bf r}}){\cal Y}_{L'}^*(\hat{{\bf r}}'), 
 \label{dstrco1}  
\end{equation}  
are related to 
each other as follows:  
\begin{equation}  
{\cal D}_{L}(\sigma,{\bf k}_\parallel)= D_{L}(\sigma,{\bf k}_\parallel) 
+ i CA^{1/2} \delta_{L0},  
\label{dldm} 
\end{equation}  
\begin{equation}  
g_{LL'}(\sigma,{\bf k}_\parallel)
= A_{LL'}(\sigma,{\bf k}_\parallel) + i A C\, \delta_{LL'}.  
\label{glal} 
\end{equation}  
$A$ and $C$ here are the familiar  numerical constants 
which have been defined by Eqs. (\ref{Adef}) and 
(\ref{CArel}), respectively.  
Relations (\ref{dldm}) and (\ref{glal}) follow 
easily from 
\begin{eqnarray}  
i\, \mbox{Im} \, G_0^+(\sigma,{\bf R})=-i C {\cal J}_0(\sigma R)   
&=& -i CA^{1/2} {\cal J}_0(\sigma R){\cal Y}_{0}(\hat{{\bf R}})\nonumber\\  
&=& \sum_{LL'}   g^0_{LL'}{\cal J}_l(\sigma r) {\cal J}_{l'}(\sigma r')  
 {\cal Y}_{L}(\hat{{\bf r}}) {\cal Y}_{L'}^*(\hat{{\bf r}'}),   
\label{equa}
\end{eqnarray}  
where
\begin{equation}  
g^0_{LL'}= -i A C\, \delta_{LL'}. 
\label{g0coef}  
\end{equation}  
In going from the first to the second equality in (\ref{equa}) 
we have used that in any space dimension ${\cal Y}_{0}(\hat{{\bf R}})=A^{-1/2}$ 
[Eq. (\ref{yps0})]. The final expressions then readily follows from 
the partial wave expansion of the free Green's function  
(see Eq. (\ref{gfpwex}) of Appendix \ref{app:fsgf}).

\newpage

\begin{center}
{\large\bf Figure captions}
\end{center}

\vspace*{2cm}

\noindent {\bf Figure 1 -}
Geometry and parameters -
A plane wave with wave vector {\bf k} incident
on $\Lambda$ with an incidence angle $\theta$.

\noindent {\bf Figure 2 -}{\bf a)} Deformation
of the integration contour for $K_{\perp n}^2>0$ 
in Eqs. (\ref{31grfr}), (\ref{32grfr}), (\ref{21grfr})
before Jordan's lemma 
is applied to the quarter-circles. {\bf b)} 
For $K_{\perp n}^2<0$ the integration contour in 
Eqs. (\ref{31grfr}), (\ref{32grfr}), (\ref{21grfr})
can be considered as a sum of two contours, one from $\eta$ to zero and 
the other from zero to $+\infty$.

\noindent {\bf Figure 3 -} Integration contour 
in the Schl\"{a}fli integral representation of 
$H_\nu^{(1)}$.

\noindent {\bf Figure 4 -} Rectilinear integration contour 
used in the derivation of the Schl\"{a}fli integral representation of 
$H_\nu^{(1)}$.

\noindent {\bf Figure 5 -} Integration contour 
in the Ewald integral representation of 
$H_\nu^{(1)}$.

\newpage

\newpage

\newpage

\newpage

\newpage

\end{document}